\documentclass[a4paper,fleqn,usenatbib]{mnras}
\usepackage[T1]{fontenc}
\usepackage{ae,aecompl}
\usepackage{graphicx}    
\usepackage{amsmath}    
\usepackage{amssymb}    
\usepackage{natbib}

\usepackage{rotating}
\usepackage{array}
\usepackage{caption}

\title[Effect of AGN on the morphological properties]{Effect of AGN on the morphological properties of their host galaxies in the local Universe}
\author[Getachew-Woreta et al.]{
Tilahun Getachew-Woreta$^{1, 2, 3}$\thanks{E-mail:tilahun85@gmail.com},
Mirjana Povi\'c$^{1, 4}$,
Josefa Masegosa$^{4}$,
Jaime Perea$^{4}$, \newauthor
Zeleke Beyoro-Amado$^{1, 2, 5}$, and
Isabel M\'arquez$^4$ 
\\
$^{1}$Astronomy and Astrophysics Research and Development Department, Entoto Observatory and Research Center (EORC), Ethiopian Space\\ 
Science and Technology Institute (ESSTI), P.O.Box 33679, Addis Ababa, Ethiopia\\
$^{2}$Addis Ababa University (AAU), P.O.Box 1176, Addis Ababa, Ethiopia\\
$^{3}$Physics Department, Bule Hora University (BHU), P.O.Box 144, Bule Hora, Ethiopia\\
$^{4}$Instituto de Astrof\'isica de Andaluc\'ia (IAA-CSIC), 18008, Granada, Spain\\
$^5$Physics Department, Kotebe Metropolitan University (KMU), P.O.Box 31248, Addis Ababa, Ethiopia
}
\date{Accepted XXX. Received YYY; in original form ZZZ}
\pubyear{2021}

\begin{document}
\label{firstpage}
\pagerange{\pageref{firstpage}--\pageref{lastpage}}
\maketitle
\begin{abstract}
The morphological classification of active galaxies may be affected by the
presence of active galactic nuclei (AGN). In this paper, we provide the
most detailed analysis on how different AGN contributions, from 5\% to 75\%, to the total optical light may affect six commonly used morphological parameters and the final classification of AGN host galaxies at z\,$\sim$\,0. We used a local sample of $>$\,2000 visually classified non-active galaxies, to quantify how the contribution of a bright nuclear point source of different intensity could affect morphological parameters such as: asymmetry, Abraham concentration index, Gini, M20 moment of light, smoothness, and Concelice-Bershady concentration index. We found that most of the morphological parameters are affected by AGN contributions above 25\%, with
late-type galaxies being more affected than early-types. We found that Gini, Abraham
concentration index and asymmetry are the most stable parameters even for AGN contributions above 25\%, in comparison to Concelice-Bershady concentration index and M20 moment of light. Smoothness parameter shall be avoided when classifying AGN host galaxies, or at least it shall be used simultaneously in combination with several other parameters.

\end{abstract}
\begin{keywords}
Galaxies; active galactic nuclei; morphological parameters; morphological classification.
\end{keywords}

\section{Introduction}
Morphology is a key parameter used to study the properties of galaxies hosting Active
Galactic Nuclei (AGN), their connection with AGN, and
their formation and evolution \citep[e.g.,][]{Kauffmann2003, Pierce2010, Povic2012, Leslie2016, Dubois2016, Gu2018}. Morphology is one of
the fundamental parameters for understanding some of the open
questions related to the effect of AGN on their host galaxy and
vice-versa, the origin of accretion material, the triggering mechanisms
that initiate an active phase in a galaxy, the duration of nuclear active phase,
etc. Therefore, the study of the morphology of active galaxies can place
important constraints on models of supermassive black hole (SMBH)
formation and growth, as well as on the formation and evolution of
galaxies across the cosmic time \citep[e.g.,][]{Mahmood2005,
  Dubois2016, Fontanot2020, Setoguchi2021}. It has been suggested that
AGN may play an important role in the morphological transformation of
galaxies from late- to early-types, being responsible for
quenching \citep[e.g.,][]{Nandra2007, Povic2012, Leslie2016,
  Dubois2016, Fontanot2020}, but also enhancing
\citep[e.g.,][]{Mahoro2017, Mahoro2019} star formation. The dependence
of the environment on AGN host galaxies in combination with morphology has
been extensively studied to better understand the AGN triggering
mechanisms. It has been suggested that minor and major
mergers do not \citep[e.g.,][and references therein]{Sharma2021,
  Silva2021} and do \citep[e.g.,][]{Cisternas2011, Ellison2019,
  Kim2020, GaoF2020} contribute significantly to the nuclear activity
in galaxies. In particular, the interpretation of morphology in terms of galaxy evolution still remains a problem. Since the determination
of the morphology depends strongly on image resolution, morphological
classification in deep surveys remains difficult, especially when
dealing with faint and high redshift sources \citep{Povic2015}.

Different methods have been used to classify galaxies in
terms of their morphology. These include visual classification
\citep[e.g.,][]{Hubble1926, Vaucouleurs1959, Sandage1996, Buta2013,
  Lintott2008, Nair2010, Simmons2017, Willett2017, Zy2020} and parametric
methods that assume an analytic model for fitting galaxy surface brightness based on its
physical and mathematical parameters \citep[e.g.,][]{Peng2010,
  Simard2011, Bar2012, Jimenez2012}. Non-parametric methods have been also commonly used, by
measuring different galaxy properties that correlate with
morphological types, such as colour, light distribution, galaxy shape,
etc. \citep[e.g.,][]{Abraham1994, Abraham1996, Conselice2000,
  Conselice2003, Huertas2008, Tasca2009, Povic2013, Mahoro2019}, being
very often mixed with the machine and deep learning for giving galaxy a probability to
be early- or late-type \citep[e.g.,][]{Huertas2008, Povic2012,
  Povic2013, Povic2015, Pintos2016, BA2019}, and/or purely based
machine and deep learning methods \citep[e.g.,][]{Kuminski2014, Huertas2015,
  Aniyan2017, Dominguez2018, Lukic2018, deDiego2020}. Each method has
its advantages and disadvantages. Visual and parametric methods are
more used for nearby and well-resolved galaxies, providing a more
detailed information about galaxy structure, while non-parametric and
machine/deep learning methods are less time-consuming and showed to be useful when dealing with large datasets, and higher redshift
sources.

There are still many inconsistencies between the results
obtained regarding the morphology of active galaxies
\citep[e.g.,][]{Kauffmann2003, Pierce2007, Georgakakis2008, Choi2009,
  Gabor2009, Povic2009, Povic2012, Mahoro2019, ramonperez2019}, and
how AGN may affect the morphological classification of their host galaxies
\citep[e.g.,][]{Gabor2009, Cardamone2010, Pierce2010}. Many previous
studies found that AGN host galaxies usually reside in spheroids or
bulge-dominated galaxies \citep[e.g,][]{Kauffmann2003, Pierce2007,
  Pierce2010, Povic2009, Povic2012, Dubois2016}, although some studies
find a higher concentration of later-types \citep[e.g.,][]{Choi2009,
  Gabor2009, Mahoro2019}. It has been also shown that morphological
properties of AGN host galaxies depend on the type of detection, where
most radio detected AGN are hosted by red sequence galaxies
\citep[e.g.,][]{Hickox2009}, X-ray detected AGN hosts occupy all areas of
the colour-magnitude diagram, but mostly green valley
\citep[e.g.,][]{Nandra2007, Georgakakis2008, Hickox2009, Povic2009,
  Povic2012}, while many of the infrared detected AGN hosts reside in slightly
bluer galaxies and occupy mainly the blue cloud region with a
significant fraction of peculiar galaxies, with clear signs of
interactions and mergers \citep[e.g.,][]{Hickox2009, GarciaBernete2015, Mahoro2019}. Although several works commented how the AGN contribution
may affect the morphological classification of their host galaxies,
finding insignificant contributions in most cases
\citep[e.g.,][]{Gabor2009, Cardamone2010, Pierce2010, Trump2015,
  Wang2017}, we are still lacking detailed analysis of how different
morphological parameters, commonly used in galaxy classification, are affected by the AGN contribution and therefore how the final
classification of galaxies will be affected.

In this work, we go a step further and study how different AGN contributions may affect the six
morphological parameters that are commonly used in non-parametric
methods for galaxy classification. We carried out all the analysis by
using a sample of a number of local, non-active galaxies with known visual
morphological classification \citep{Nair2010}. In this paper, we
present the results obtained for the original sample at
z\,$\sim$\,0. The simulations at fainter magnitudes
and higher redshifts will be presented in a forthcoming paper
(Getachew et al., in prep.).
	
The paper is organized as follows: in section \ref{sec2} we describe
the used data, along with a brief description of the generated source
catalogue. The applied methodology is described in section \ref{sec3},
while all analysis are given in section \ref{sec4}. Section
\ref{sec5} gives the results and discussion of our work, and finally,
we draw our conclusions in section \ref{sec6}. We assumed cosmological
model with $\Omega_{\Lambda}$=0.7, $\Omega_{M} = 0.3$, and $H_{0}$ =
70 kms$^{-1}$Mpc$^{-1}$. All magnitudes in this paper are given in the
$AB$ system \citep{Oke2015}.
 
\section{Data}\label{sec2}
\begin{figure*}
\centering
\includegraphics[width=2.995in,height=1.93in]{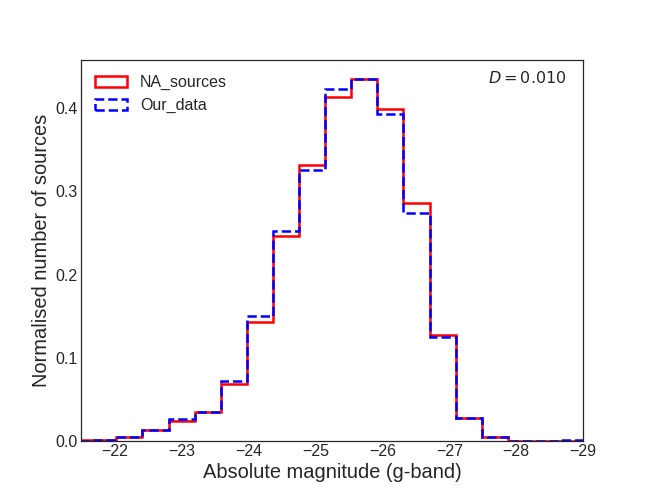}
\includegraphics[width=2.995in,height=1.93in]{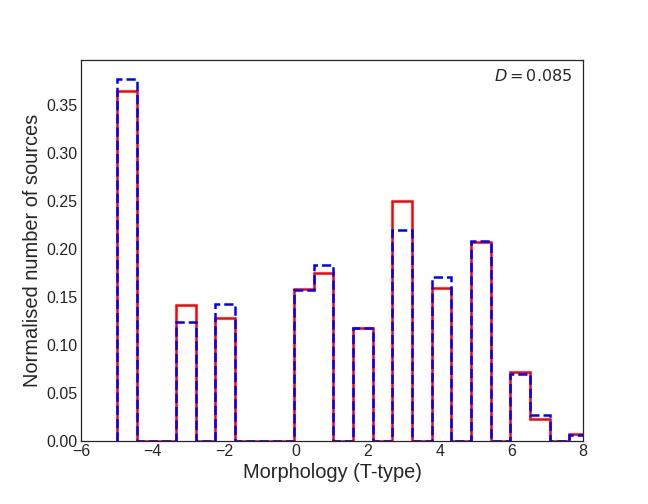}
\includegraphics[width=2.995in,height=1.93in]{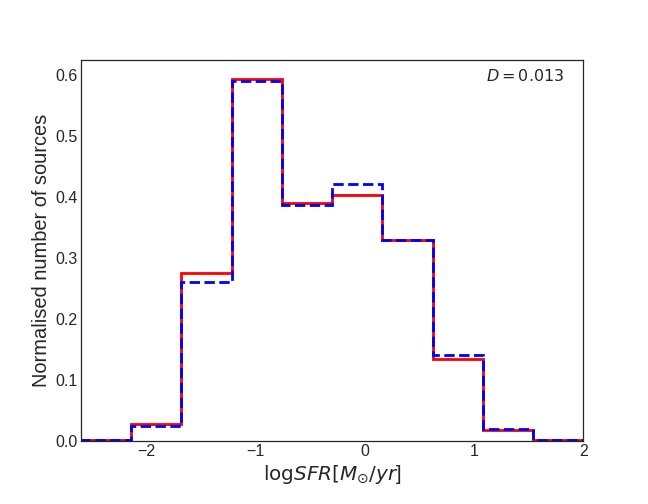}
\includegraphics[width=2.995in,height=1.93in]{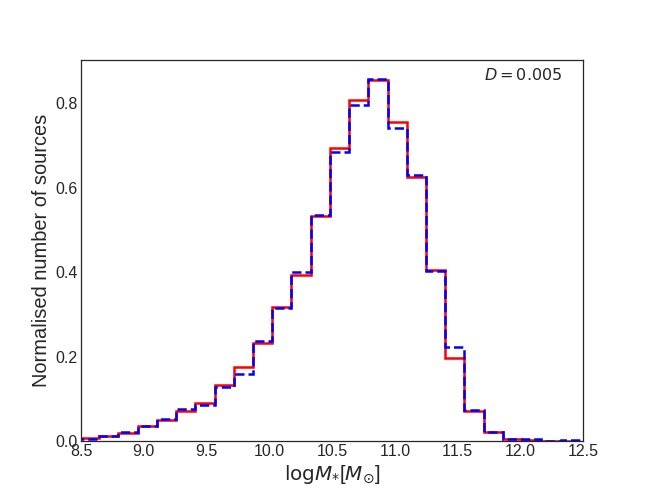}
\includegraphics[width=2.995in,height=1.93in]{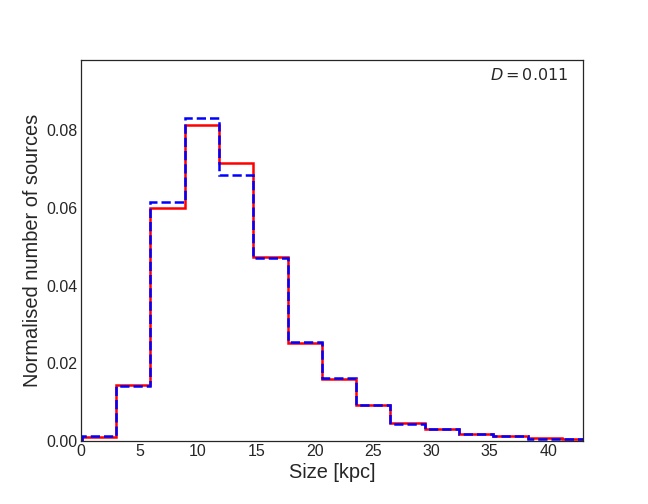}
\caption{The comparison between the properties of the full \protect\cite{Nair2010} sample (red lines) and the sub-sample of 8000 galaxies (blue lines) used in this work, including absolute magnitude in g-band (top left), morphology (top right), SFR (middle to left), stellar mass (middle to right), and size (bottom). The KS test maximum
deviation D parameter between the two distributions is given in all cases.}.
\label{fig1}
\end{figure*}

To study the effect of the AGN contribution on the morphological parameters
of their host galaxies in the local Universe, we used an initial sample of 8000
local galaxies visually classified in the g-band by \cite{Nair2010}. The authors used
the data from the Sloan Digital Sky
Survey\footnote{https://www.sdss.org/} (SDSS; \citealt{York2000},
\citealt{Stoughton2002}) Data Release 4 (DR4) down to an
extinction-corrected apparent magnitude of $g<16$ and in the redshift
range of $0.01 \leq z \leq 0.1$ (with a mean redshift of $0.04$). The
galaxies were selected randomly out of $\sim 14000$ sources contained
in the \cite{Nair2010} catalogue, as explained in detail in
\citet{Povic2013}. To make sure that the selected sub-sample of local
galaxies is a fair representation of the parent sample, we compared the
two samples in terms of their general properties. In Fig.~\ref{fig1} we
show the distributions for both samples for the absolute magnitude in
g-band, morphology, star formation
rate (SFR), and stellar mass, where the last two parameters have been
obtained from the
MPA-JHU\footnote{https://wwwmpa.mpa-garching.mpg.de/SDSS/} SDSS
catalogue \citep{Brinchmann2004, Salim2007}. The Kolmogorov-Smirnov
(KS) test for all comparison plots indicates that the maximum
deviation (D) between the two distributions is $\sim$\,0, indicating
that the two samples are consistent.

\begin{figure}
\begin{center}
\includegraphics[height= 2.5in, width=3.4in]{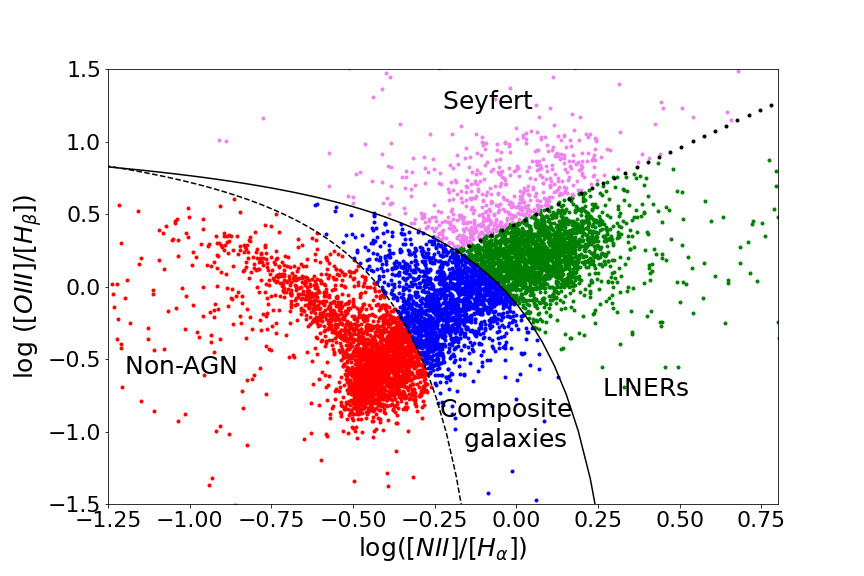}
\caption{The diagnostic BPT-NII diagram separating the selected sample of local galaxies into: star-forming (red symbols), composite (in blue), Seyfert 2 (in indigo) and LINERs (in green). Used separation limits between four classes (from left-to-right) correspond to: \protect\citet[dashed line]{Kauffmann2003}, \protect\citet[solid line]{Kewley2006}, and \protect\citet[dotted line]{Schawinski2007}.} 
\label{BPT1}
\end{center}
\end{figure}

Taking into account the nature of this work, we need a sample of
non-active galaxies for our analysis. To eliminate AGN hosts, we
cross-matched our sample of 8000 galaxies with the MPA-JHU SDSS DR7
emission-line catalogue of $\sim$\,920000 sources, using a
cross-matching radius of 2\,arcsec. We obtained a total of 7940
counterparts with available redshifts and emission lines with
signal-to-noise ratio S/N\,$>$\,3. For these galaxies, we plotted the
\cite{Baldwin1981} BPT-NII diagram as shown in Fig.~\ref{BPT1}, using
the emission-line ratios of $[NII]\lambda 6584/H\alpha $ and
$[OIII]\lambda 5007/H\beta$. To separate star-forming (non-AGN)
galaxies from composite and AGN galaxies we used the separation limits
given by \cite{Kauffmann2003} and \cite{Kewley2006}. We also
checked the number of Seyfert 2 and LINER sources using the separation
limit of \cite{Schawinski2007}. We found in total 2744 (35\%) 
star-forming galaxies, 1918 (24\%) composites, 594 (7\%) Seyfert
2s, and 2684 (34\%) LINERs.

\begin{figure}
\begin{center}
\includegraphics[width=\columnwidth]{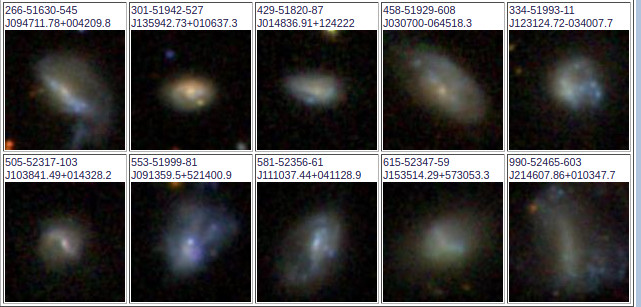}
\caption{An example of peculiar galaxies analysed in this paper, with clear signs of disturbed structures.} 
\label{fig_peculiar}
\end{center}
\end{figure}

After selecting these 2744 non-AGN galaxies, we went through the visual
inspection of their images and we further removed those galaxies with
superposed foreground bright stars, and also galaxies with
signs of interactions and mergers. Our final sample of non-active
galaxies contains 2301 sources. Using the visual morphological
classification of \cite{Nair2010} and their T-type classification
scheme (see their Table 1), we have 471 early-type galaxies
(elliptical and lenticular, with T-type\,$\le$\,0), 891 early-spirals
(0\,$>$\,T-type\,$\le$\,4), 889 late-spirals (4\,$>$\,T-type\,$\le$\,8), and 50 irregular and peculiar galaxies (T-type\,$>$\,8). We inspected visually the images of those galaxies with T-type\,$>$\,8 and found that the majority ($> 80\%$)
of them are not irregular galaxies, but rather disturbed galaxies with clear signs of
interactions, and therefore along this paper we will call these
galaxies peculiar. An example of 10 galaxies included under this
class is given in Fig.~\ref{fig_peculiar}.

\section{Methodology}\label{sec3}
In this section, we analysed six of the most used parameters in
the non-parametric morphological classification of galaxies, and how
sensitive they are on AGN contribution when dealing with the
classification of active galaxies. We measured the six parameters for the
sample of 2744 non-AGN galaxies. We then
artificially added in their center different AGN
contributions (from 5\%\,-\,75\%), and we again measured the same parameters in a
consistent way.

\subsection{Morphological parameters of the local sample of non-active galaxies}\label{sec31}

The six morphological parameters analysed in this work have been used in
many previous studies to distinguish between early- (elliptical and
lenticular) and late-type (spiral and irregular) galaxies
\citep[e.g.,][]{Sca2007, Huertas2008, Huertas2009, Povic2012,
  Povic2013, Povic2015, Licitra2016, Pintos2016, Tars2018, BA2019,
  Mahoro2019, Baes2020, Bignone2020}. These six parameters are:\\

\indent \textbf{Asymmetry index} (hereafter ASYM;
\citealt{Abraham1994}). This parameter describes the level of
asymmetry in the shape of one galaxy. It is calculated by comparing an
object flux in every pixel using an original image and the rotated one
by $180^{o}$, and taking into account the background. It can be
expressed as:

\begin{equation}
ASYM =\frac{\Sigma {|I(i,j)-I_{180}(i, j)|/2}}{I(i, j)}-\frac{\Sigma {|B(i,j)-B_{180}(i, j)|/2} }{I(i, j)},
\end{equation}

where I$(i,j)$ and I$_{180}(i,j)$ are the fluxes in the $(i,j)$ pixel
position on the original and rotated image, while B$(i,j)$ and
B$_{180}(i,j)$ are the fluxes of the background in the original and
rotated image, respectively. Highly symmetrical sources, such as
undisturbed elliptical or lenticular galaxies, will have minimal
residuals and small values of ASYM, while the significant residuals of
spiral, irregular, and peculiar galaxies will correspond to higher
values of ASYM. In general, the allowed range of this parameter is
0\,$<$\,ASYM\,$<$\,1.\\

\indent\textbf{Abraham concentration index} (hereafter CABR;
\citealt{Abraham1996}). This is one of the most used light
concentration parameters. It is measured as the ratio between the flux
at 30\% $(F_{30})$ of the Petrosian radius \citep{Petrosian1976} and
the total flux $(F_{tot})$ of the galaxy:

\begin{equation}
CABR=F_{30}/F_{tot}.
\end{equation}
The typical range of values of this parameter is $0 < CABR < 1$, where galaxies with higher light concentration in their centers have higher values of CABR.\\

\indent \textbf{Gini coefficient} (hereafter GINI;
\citealt{Abraha2003, Lotz2004}). This parameter has been adopted from
the economic statistics \citep{Gini1912} and is now used in astronomy
to describe the distribution of light among the pixels associated with
a galaxy.  If $n$ is the number of pixels assigned to a galaxy,
$f_{i}$ the flux from pixel $i$ (ordered in such a way that $f_{ i}$
increases with the pixel index), and $\bar{f}$ the mean values of the
pixel, GINI parameter can be defined as:

\begin{equation}
GINI=\frac{1}{|\bar{f}|n(n-1)}{\Sigma^{n}_{i}}{(2i-n-1)|f_{i}|}.
\end{equation}

The values of GINI cover the range $0< GINI < 1$, where low values
characterise an equal distribution of light across all galaxy pixels,
while high values point to extreme inequality in light distribution
contained within a small number of pixels (e.g., in a single bright
nucleus of a galaxy or a galaxy with multiple bright nuclei). This
parameter has been suggested to be the efficient one for morphological
measurements of strongly lensed galaxies \citep{Florian2016a}.\\

\indent \textbf{M20 moment of light} (hereafter M20;
\citealt{Lotz2004}). The value of this parameter depends on the
spatial distribution of the light in a galaxy relative to its
center, summed over the 20\% brightest pixels and normalised by the
total second-order moment of light (Mtot):

\begin{equation}
{M20=log_{10}\left({\frac{\Sigma_{i}{M_{i}}}{M_{tot}}}\right),} 
\end{equation}

while the flux $f_{i}$ in each pixel in comparison to the total flux
$f_{tot}$ is $\Sigma_{i}{f_{i}<0.2 f_{tot}}$,\\ and where

\begin{equation}
M_{tot}=\Sigma^{n}_{i}{M_{i}}=\Sigma^{n}_{i}{f_{i}[(x_{i}-x_{c})^{2}+(y_{i}-y_{c})^{2}]},
\end{equation}

with $(x_{c}, y_{c})$ standing for the galaxy center obtained by
minimising the total light \citep{Huertas2008}. This parameter in
general has values in the range $-3 < M20 < 0$. It is very sensitive to the
spatial distribution of light. For undisturbed early-type
galaxies when the brightest 20\% of the light may easily be positioned
close to the galactic center, M20 approaches values of -2.5 or -3.0. On
the other hand, the presence of multiple bright sources of light that
are spatially separated from the galaxy center (e.g., bars, spiral arms,
strong star-forming regions, etc.) increase the value of M20 toward 0
\citep{Conselice2014}.\\

\indent \textbf{Smoothness or clumpiness} (hereafter SMOOTH;
\citealt{Conselice2000}). This parameter is sensitive to the
small-scale structures in galaxies. To measure it, the original galaxy
image is smoothed with a boxcar of a given width and subtracted
from the original image. SMOOTH is then obtained as:

\begin{equation}
SMOOTH =\frac{\Sigma {|I(i,j)-I_{s}(i, j)|/2}}{I(i, j)}-\frac{\Sigma {|B(i,j)-B_{s}(i, j)|/2} }{I(i, j)},
\end{equation}

where $I(i,j)$, $B(i,j)$ are the flux of the galaxy and background,
respectively, in the pixel position $(i,j)$ of the original image,
while $I_{s}(i,j)$ and $B_{s}(i,j)$ represent the same in the case of
the smoothed image. This parameter has values $0<SMOOTH<1.0$, where higher
values indicate a larger number of small-scale structures in a galaxy.\\

\indent \textbf{Conselice-Bershady concentration index} (hereafter
CCON; \citealt{Bersh2000}). Similar to CABR and GINI, this parameter
gives the level of light concentration within a galaxy. It is
measured as the logarithm of the ratio of the circular radii containing
80\% and 20\% of the total flux:

\begin{equation}
CCON= 5log(r_{80}/r_{20}).
\end{equation}

In general, lower CCON values correspond to lower fractions of light in the central region. \\ 

We have written an in-house IDL code for measuring all of the six
described parameters of the selected sample of the non-AGN
galaxies. To obtain all needed input parameters we first run
SExtractor \citep{Bertin1996} using the SDSS g-band images. When
necessary, the galaxy centre is
determined by minimizing ASYM \citep{Abraham1996}, while the total
flux is defined as the one contained within 1.5 times the Petrosian
radius \citep{Huertas2008} as measured by SExtractor.

\subsection{Morphological parameters of the local sample with added AGN contributions}
\label{sec32}
For analysing how the presence of an AGN may affect the different morphological parameters
and hence the final morphological classification of active galaxies, we used the
selected sample of star-forming (non-AGN) galaxies and simulated AGN hosts
by adding in their centers different AGN contributions. For simulating
the AGN contribution, for each galaxy we used its corresponding point
spread function (PSF) image in the g-band obtained from the SDSS database
\citep[see][for more details]{Povic2013}. To model the PSF we used the Moffat
function \citep{Moffat1969}, which showed to be its good analytical
approximation \citep[e.g.,][]{Molina1992, Trujillo2001}. For each galaxy we
constructed five simulated images using different fractions of AGN
contribution to the total galaxy flux: 5\%, 10\%, 25\%, 50\%, and
75\%. The total galaxy flux was measured by running SExtractor
code on the g-band images. Once we obtained simulated images of galaxies
with the different AGN contributions added we repeated the steps from
Sec.\,\ref{sec31}. We first run SExtractor on all simulated images to
obtain the different input parameters needed for measuring the morphological
parameters. Secondly, for each AGN contribution of 5\%\,-\,75\% we
measured the six morphological parameters following the procedure in Sec.\,\ref{sec31}.

\begin{figure*}
\begin{center}
\includegraphics[height=2.0in, width=2.8in]{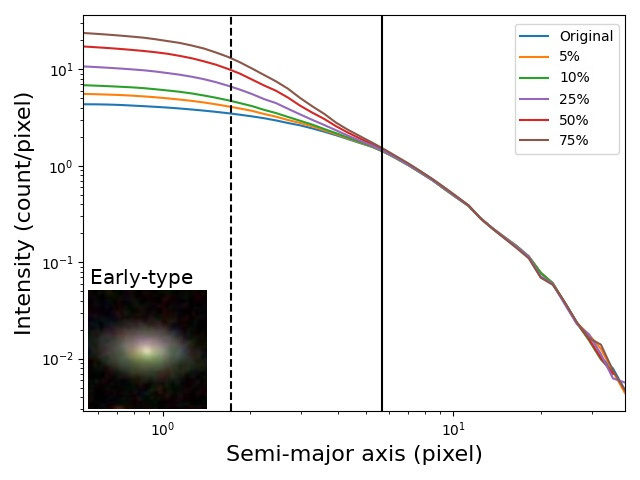}
\includegraphics[height=2.0in, width=2.8in]{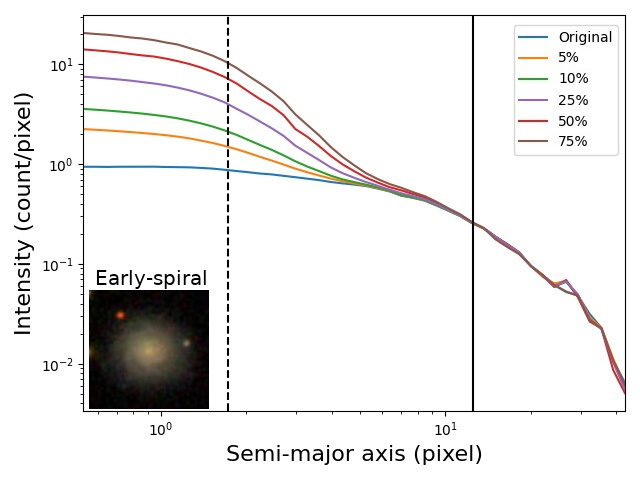}
\includegraphics[height=2.0in, width=2.8in]{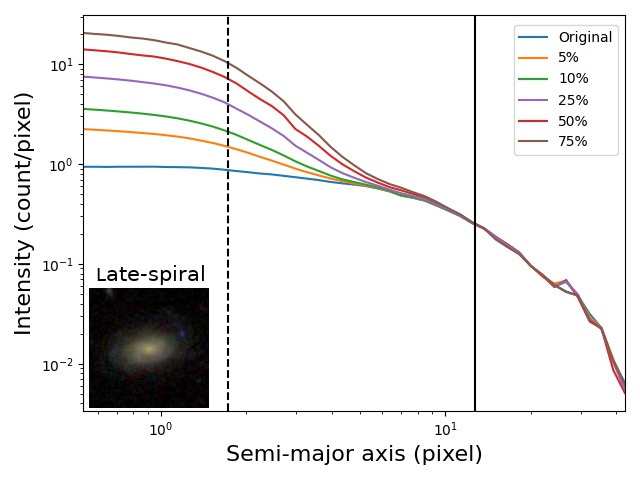}
\includegraphics[height=2.0in, width=2.8in]{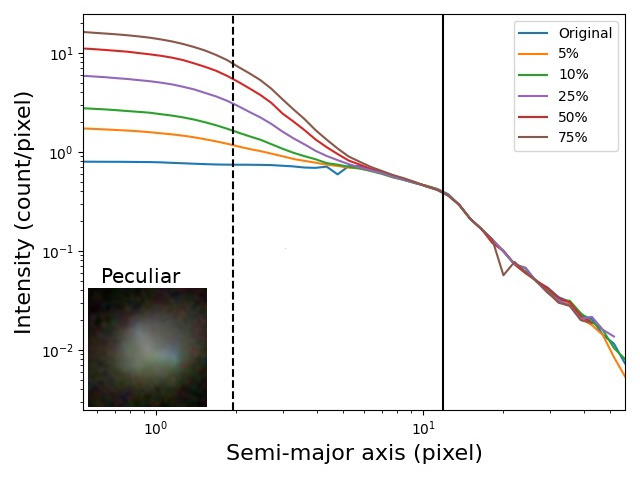}
\caption{The example of surface brightness profiles of early-type (top left), early-spiral (top-right), late-spiral (bottom left), and peculiar (bottom right) galaxy when using the original g-band image (blue line) and simulated once with AGN contributions of 5\% (yellow line), 10\%  (green line), 25\% (purple line), 50\% (red line), and  75\% (brown line). The black dashed and solid vertical lines indicate the PSF and galaxy Petrosian half-light radius.}
\label{profile1}
\end{center}
\end{figure*}
	
\begin{table*}
\caption{Additional information of galaxies showed in Fig.\,\ref{profile1}, including: object ID, right ascension and declination, Petrosian radius at 50\% and 90\% of the total galaxy flux, total galaxy flux, morphology, Petrosian radius at 50\% and 90\% of the PSF flux, and PSF flux, respectively.}
\begin{tabular} {c|c|c|c|c|c|c|c|c|c|c}
\hline
Object ID&RA&DEC& $R_{50}$\%    & $R_{90}$\%   & Total-flux & Morphology& PSF $R_{50}$\%& PSF $R_{90}$\%&PSF Total-flux\\ 
&(deg)&(deg)&(pix)&(pix)&(counts)& &(pix)&(pix)&(counts) \\ \hline
535-51999-573&218.32646&2.7014122 &5.69 &14.30 & 387.95 & early-type  & 1.73& 3.88  & 347.51    \\ \hline
420-51871-506&13.712063&15.278116&12.51 &26.31 & 407.49 & early-spiral&1.74    & 3.76  & 357.42   \\ \hline
476-52314-66&144.95767 &-0.56153035&17.74 &27.96 & 362.99 & late-spiral& 1.71   &  3.90 &  326.41  \\ \hline
615-52347-59& 233.80957 & 57.514824 &11.90&26.03&382.77&  peculiar& 1.95   &  4.13 &  432.98\\ \hline
\end{tabular}                                                                       
\label{table:1}
\end{table*}

Figure\,\ref{profile1} shows the surface brightness profiles using the original image (blue line) and
the simulated images with 5\% to 75\% of AGN contribution, in the case of
an early-type galaxy (top left), an early-spiral (top right), a late-spiral
(bottom left), and a peculiar galaxy (bottom right). The two vertical
lines in each plot correspond to the Petrosian half-light radius
(R$_{50}$\%) of the PSF (dashed line) and galaxy (solid line). The
size of each galaxy at 50\% and 90\%, its total flux and
corresponding PSF values are given in Table\,\ref{table:1} as an
example.

Finally, as a result of the work
described in Sec.\,\ref{sec31} and here, for each galaxy and each
morphological parameter we obtained six measurements: the real one
using original images, and simulated ones when having 5\% to 75\% of
AGN contribution in comparison to the total galaxy flux.

\section{Analysis and results}\label{sec4}
In this section, we measure the effect of the AGN contribution on the morphological
parameters when considering the total sample of selected galaxies and when taking into account their morphological type. We also test the
effect of the AGN on some of the most used morphological diagrams when
dealing with the classification of galaxies.

\subsection{Effect of the AGN on the morphological parameters of the total sample}\label{sec41}
Using the total sample of the 2301 selected sources described in Sec.~\ref{sec2},
we analysed in detail the effect of AGN on the six morphological
parameters described in Sec.~\ref{sec31}. For each parameter obtained
using original and simulated images (with 5 different added AGN
contributions), we measured their median values, first quarter
(Q1\footnote{Q1 corresponds to 25\% of sample.}), and third quarter
(Q3\footnote{Q3 corresponds to 75\% of sample.}), where the range
Q1-Q3 corresponds to 50\% of the sample. We used the median values of samples plus the change in the [Q1-Q3] range as indicators of the change (or scatter) in samples with AGN contribution.

\begin{figure}
\begin{center}
\includegraphics[width=\columnwidth]{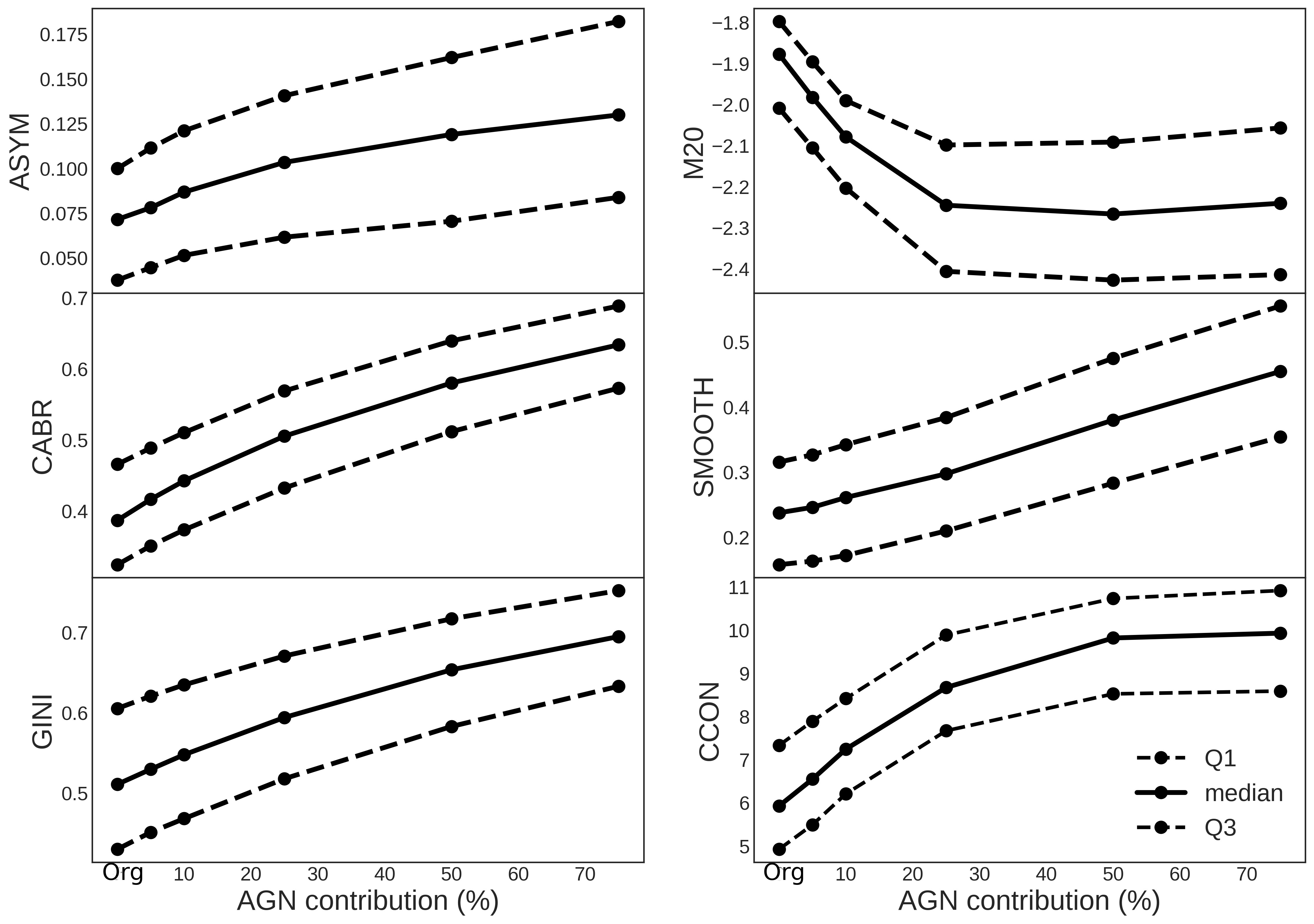}
\caption{Distribution of median values (black solid lines) of ASYM (top left), CABR (middle left), GINI (bottom left), M20 (top right), SMOOTH (middle right) and CCON (bottom right) in relation to AGN contribution to the total flux from 0\% up to 75\%. Dashed lines correspond to Q1-Q3 range where 50\% of the sample is located.}
\label{sec:figlocal}
\end{center}
\end{figure}

Figure \ref{sec:figlocal} shows the effect of the AGN contribution
on the six morphological parameters used in this work: ASYM (top left),
CABR (middle left), GINI (bottom left), M20 (top right), SMOOTH
(middle right), and CCON (bottom right). We represent how the median values
of these parameters (solid lines) change from the values measured on the original images
(without AGN being added, indicated in Fig.~\ref{sec:figlocal} as
"Org") when 5\%, 10\%, 25\%, 50\% and 75\% of AGN contribution is added.
We also represent the corresponding Q1 and Q3 measurements
(dashed lines), that can be considered as an indicator of the error.

Most of the parameters change with the added AGN contribution in comparison to
real values, with gradually larger differences being observed as the
AGN contribution to the total flux increases. In Table~\ref{table_quantification}, the effect of the AGN has been measured using the median values of
each parameter and each sample (total and for different morphological types), measured on original images (param\_orig) and after adding different AGN contributions
(param\_AGN), such as: (param\_AGN\,-\,param\_orig)/param\_orig. A total number of sources with measured parameters under each sample is also listed in the table. CCON is one of the most sensitive parameters to AGN contribution, facing 22\% and 46\% of
change (toward higher values) in case of 10\% and 25\% of added AGN contribution, respectively. On the other side, ASYM is one of the parameters that almost doesn't change independently on AGN contribution. As can be seen in Fig.~\ref{sec:figlocal} it takes median values between 0 and 0.13, and therefore we are not listing any values for this parameter in Table~\ref{table_quantification}. This parameter has been measured for 72\%, 91\%, 71\%, 61\%, and 66\% of the total sample, early-type, early-spiral, late-spiral, and peculiar galaxies, respectively. After ASYM, GINI is the least affected parameter
changing only 8\%, 16\%, 27\%, and 37\% for 10\% 25\%,
50\%, and 75\% of AGN contribution, respectively. Up to 10\% of AGN, CABR and M20 show 10-15\% of change. Similar to CABR and M20, SMOOTH is also
showing a low level of contamination below 25\% of AGN contribution, however, this parameter is very sensitive to
noise, as shown previously \citep[e.g.,][]{Huertas2008, Povic2013, Povic2015}, and it could be measured properly for a very small
sub-sample (44\%) of the total number of galaxies (see Table~\ref{table_quantification}). In general, we can observe that AGN
contribution to the total flux of 25\% and above produce significant changes in the parameters of $\gtrsim$\,30\% (or $\gtrsim$\,20\% in the case of GINI and M20), except in the case of ASYM.

\begin{table*}
\caption{Quantification of the effect of 5\%\,-\,75\% of AGN on the five morphological parameters in the case of the total sample, early-type, spiral early-type, spiral late-type, and peculiar galaxies. ASYM parameter doesn't vary much with AGN contribution (see Fig.~\ref{sec:figlocal} and the text), and therefore, it is not listed in this table. Fraction of galaxies with good measurements of analysed morphological parameters of each sample is given in column 2, and in the text in the case of ASYM.}
\begin{tabular} {|c|c|c|c|c|c|c|}
\hline
\bf{Measured parameter}&\bf{Number of sources}&\bf{5\% }&\bf{10\% }&\bf{25\% }&\bf{ 50\% }&\bf{ 75\% }\\ 
\bf{(sample)}&&\bf{AGN}&\bf{AGN}&\bf{AGN}&\bf{AGN}&\bf{AGN}\\ \hline
\bf{CABR (total sample)}&2292 (100\%) &8\%\,$\pm$\,2\%   &13\%\,$\pm$\,2\%  &31\%\,$\pm$\,5\%   &49\%\,$\pm$\,7\%  &60\%\,$\pm$\,8\% \\ \hline
\bf{CABR (early-type)}&470 (100\%)  &4\%\,$\pm$\,1\%    &7\%\,$\pm$\,1\%   &18\%\,$\pm$\,2\%   &27\%\,$\pm$\,3\%   &35\%\,$\pm$\,4\%  \\ \hline
\bf{CABR (early-spiral)}&887 (100\%)  &7\%\,$\pm$\,1\%  &13\%\,$\pm$\,2\% &30\%\,$\pm$\,3\% &49\%\,$\pm$\,6\% &66\%\,$\pm$\,9\%   \\ \hline
\bf{CABR (late-spiral)}&885 (100\%)  &9\%\,$\pm$\,2\%&18\%\,$\pm$\,3\% &35\%\,$\pm$\,2\%   &55\%\,$\pm$\,6\% &73\%\,$\pm$\,9\%   \\ \hline
\bf{CABR (peculiar)}&50 (100\%)  &3\%\,$\pm$\,2\% &8\%\,$\pm$\,2\%&24\%\,$\pm$\,3\% &42\%\,$\pm$\,10\%  &58\%\,$\pm$\,13\%\\ \hline
\bf{GINI (total sample)}&2294 (100\%) &4\%\,$\pm$\,1\%   &8\%\,$\pm$\,2\%  &16\%\,$\pm$\,5\% &27\%\,$\pm$\,9\% &37\%\,$\pm$\,12\% \\ \hline
\bf{GINI (early-type)}&470 (100\%)  &3\%\,$\pm$\,1\%    &5\%\,$\pm$\,1\% &11\%\,$\pm$\,3\%  &20\%\,$\pm$\,4\% &25\%\,$\pm$\,5\% \\ \hline
\bf{GINI (early-spiral)}&888 (100\%)&4\%\,$\pm$\,1\%  &8\%\,$\pm$\,2\% &16\%\,$\pm$\,5\% &27\%\,$\pm$\,9\%  &37\%\,$\pm$\,11\%  \\ \hline
\bf{GINI (late-spiral)}&886 (100\%)  &5\%\,$\pm$\,2\%   &9\%\,$\pm$\,2\%  &18\%\,$\pm$\,4\%    &33\%\,$\pm$\,7\%   &42\%\,$\pm$\,9\%     \\ \hline
\bf{GINI (peculiar)}&50 (100\%)&2\%\,$\pm$\,1\%    &6\%\,$\pm$\,2\% &12\%\,$\pm$\,2\%     &20\%\,$\pm$\,3\%   &30\%\,$\pm$\,4\%    \\ \hline
\bf{M20 (total sample)} &2185 (95\%)       &6\%\,$\pm$\,1\%    &11\%\,$\pm$\,1\% &20\%\,$\pm$\,2\%     &21\%\,$\pm$\,3\%   &19\%\,$\pm$\,3\%     \\ \hline
\bf{M20 (early-type)}&447 (96\%) &2\%\,$\pm$\,1\%    &4\%\,$\pm$\,1\%  &6\%\,$\pm$\,1\%     &4\%\,$\pm$\,1\%    &2\%\,$\pm$\,2\%      \\ \hline
\bf{M20 (early-spiral)}&846 (95\%)     &5\%\,$\pm$\,2\%    &11\%\,$\pm$\,2\%  &19\%\,$\pm$\,2\%    &21\%\,$\pm$\,2\%  &20\%\,$\pm$\,2\%     \\ \hline
\bf{M20 (late-spiral)}&836 (94\%)      &14\%\,$\pm$\,2\%  &27\%\,$\pm$\,4\% &44\%\,$\pm$\,7\%    &47\%\,$\pm$\,5\%  &46\%\,$\pm$\,5\%    \\ \hline
\bf{M20 (peculiar)}&48 (96\%) &3\%\,$\pm$\,1\%   &7\%\,$\pm$\,1\%  &19\%\,$\pm$\,3\%     &18\%\,$\pm$\,2\%  &17\%\,$\pm$\,4\% \\ \hline
\bf{SMOOTH (total sample)}&1023 (44\%)   &4\%\,$\pm$\,1\%    &8\%\,$\pm$\,1\%  &29\%\,$\pm$\,7\%     &58\%\,$\pm$\,15\%   &87\%\,$\pm$\,24\% \\ \hline
\bf{SMOOTH (early-type)}&352 (75\%)      &6\%\,$\pm$\,5\%    &16\%\,$\pm$\,6\% &32\%\,$\pm$\,15\%     &64\%\,$\pm$\,29\%   &84\%\,$\pm$\,33\%  \\ \hline
\bf{SMOOTH (early-spiral)}&347 (39\%)    &6\%\,$\pm$\,4\%    &12\%\,$\pm$\,4\% &24\%\,$\pm$\,6\%    &58\%\,$\pm$\,5\%  &87\%\,$\pm$\,13\%     \\ \hline
\bf{SMOOTH (late-spiral)}&291 (33\%)   &3\%\,$\pm$\,2\%    &9\%\,$\pm$\,3\% &26\%\,$\pm$\,2\%     &78\%\,$\pm$\,8\%  &113\%\,$\pm$\,9\%    \\ \hline
\bf{SMOOTH (peculiar)}&30 (60\%)       &8\%\,$\pm$\,4\%   &13\%\,$\pm$\,4\% &30\%\,$\pm$\,5\%     &65\%\,$\pm$\,8\%   &113\%\,$\pm$\,29\%    \\ \hline
\bf{CCON (total sample)}&2186 (95\%)      &10\%\,$\pm$\,2\%   &22\%\,$\pm$\,6\% &46\%\,$\pm$\,11\%     &66\%\,$\pm$\,14\%   &67\%\,$\pm$\,13\%  \\ \hline
\bf{CCON (early-type)}&457 (97\%) &5\%\,$\pm$\,1\%   &9\%\,$\pm$\,1\%  &17\%\,$\pm$\,2\%    &18\%\,$\pm$\,2\%   &18\%\,$\pm$1\%   \\ \hline
\bf{CCON (early-spiral)}&855 (96\%)      &10\%\,$\pm$\,1\%   &22\%\,$\pm$\,3\% &53\%\,$\pm$\,6\%    &72\%\,$\pm$\,7\%  &73\%\,$\pm$\,6\% \\ \hline
\bf{CCON (late-spiral)}&836 (94\%)      &11\%\,$\pm$\,3\%  &23\%\,$\pm$\,3\% &50\%\,$\pm$\,5\%    &76\%\,$\pm$\,8\%  &78\%\,$\pm$\,9\% \\ \hline
\bf{CCON (peculiar)}&49 (98\%)  &10\%\,$\pm$\,3\%  &19\%\,$\pm$\,4\% &35\%\,$\pm$\,5\%     &51\%\,$\pm$\,5\%  &68\%\,$\pm$\,8\%   \\ \hline
\end{tabular}                                                                              
\label{table_quantification}
\end{table*}

\begin{figure}
\begin{center}
\includegraphics[width=\columnwidth]{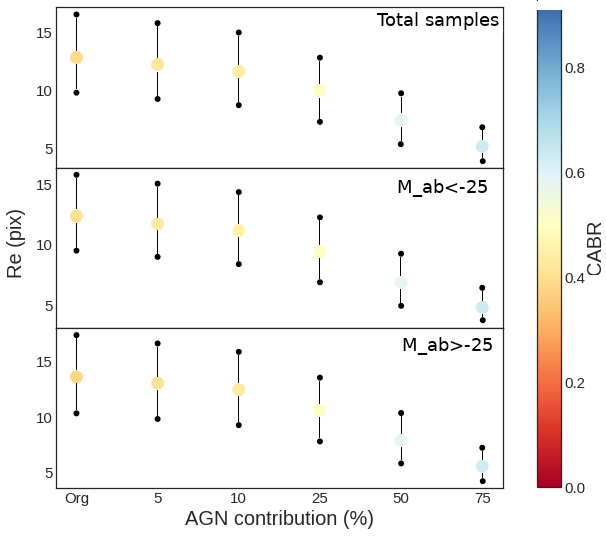}
\caption{The effective radius (Re) as a function of AGN contribution for a total sample (top), and sources with the g-band absolute magnitude of Mab\,$<$\,-25 (middle) and Mab\,$>$\,-25 (bottom). The colour bar shows the relative change of CABR morphological parameter as an example. The effective radius is represented through median values, while the vertical lines indicate the Q1 and Q3 quartiles.}
\label{Error_bar}
\end{center}
\end{figure}

In addition to the impact of AGN on measured morphological parameters, we also checked the effect of AGN in relation to the galaxy size. In Fig.~\ref{Error_bar}, we showed the median effective radius measured on original and simulated (with 5-75\% of AGN) images for a total sample (top panel) and for two different ranges of absolute magnitude (g-band Mab\,$<$\,-25 (middle panel) and Mab\,$>$\,-25 (bottom panel)). Just as an example, we showed also the relative change of CABR parameter. We can observe that in all three cases for a total sample and the two samples with different g-band absolute magnitude, the effective radius starts to be more affected when AGN contribution is above 25\%, indicating that smaller galaxies will be more affected. In all three cases, independently on brightness, the change in the effective radius is $<$\,10\% up to 10\% of AGN contribution and becomes of the order of 20\%, 40\%, and 60\% for larger AGN contributions of 25\%, 50\%, and 75\%. The median values of the effective radius obtained when adding 50\% and 75\% of AGN, correspond to only 7\% and 1\% of sources in the original sample, respectively. We also checked the effect of AGN on the Petrosian radius, finding insignificant effect of $<$\,4\% up to 25\% of AGN contribution, increasing to 13\% and 28\% of change for 50\% and 75\% of AGN contribution, respectively. Finally, we also checked the effect of size on morphological parameters without having an AGN. In general, we do not observe any significant effect of size on ASYM, SMOOTH, M20 (except a small effect of $<$\,20\% for ET galaxies), and CCON parameters. A slight change in CABR with size has been observed for a total sample, where more compact galaxies (with CABR\,$\sim$\,0.5) have \,$\sim$\,25\% smaller sizes in comparison to less compact galaxies (CABR\,$\sim$\,0.2\,-\,0.3). This effect is a bit stronger in the case of the GINI parameter, where more compact galaxies with GINI\,$\sim$\,0.7 have \,$\sim$\,40\% smaller sizes in comparison to less compact galaxies (GINI\,$\sim$\,0.35).

\subsection{Effect of AGN on the studied parameters for the different morphologies}\label{sec42}
In this section, we present a similar analysis as above, but now taking
into account morphology. We use the four broad morphological classes as described in
Sec.\,\ref{sec2}.

\begin{figure}
\begin{center}
\includegraphics[width=\columnwidth]{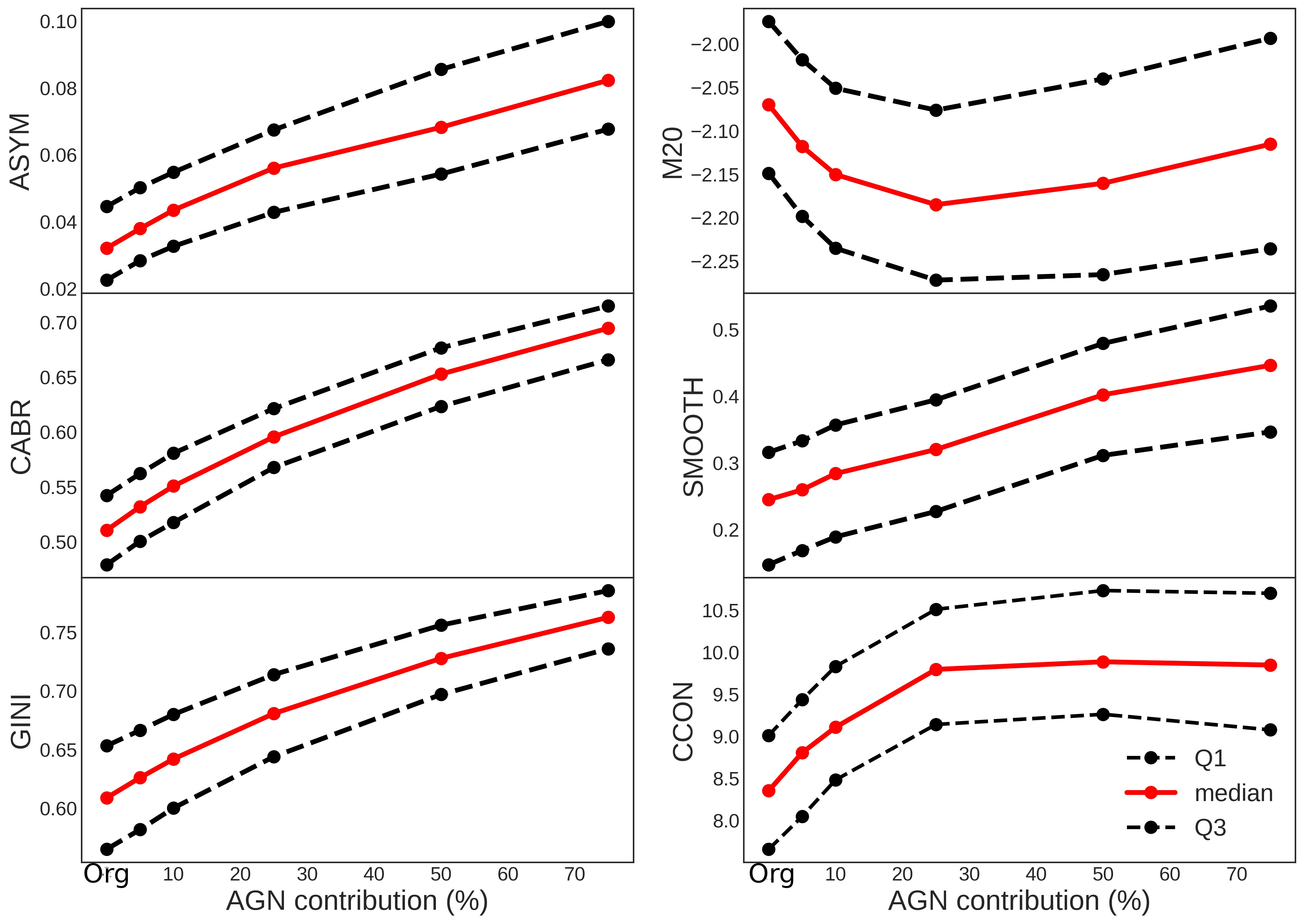}
\caption{Same as Fig. \ref{sec:figlocal}, but only for early-type galaxies.}
\label{sec:Ell}
\end{center}
\end{figure}

\begin{figure}
\begin{center}
\includegraphics[width=\columnwidth]{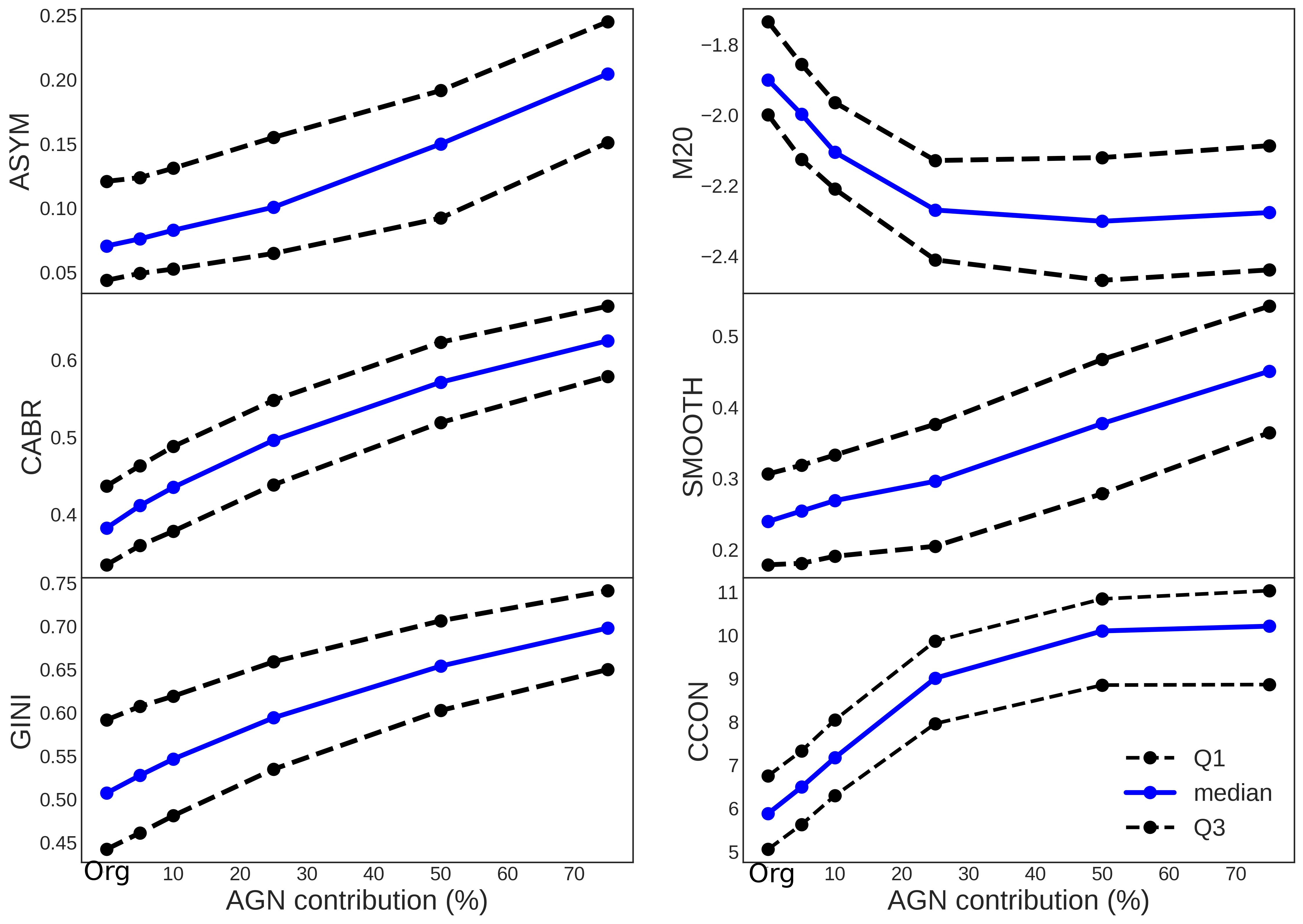}
\caption{Same as Fig. \ref{sec:figlocal}, but only for early-spiral galaxies.}
\label{sec:AGNspe}
\end{center}
\end{figure}

\begin{figure}
\begin{center}
\includegraphics[width=\columnwidth]{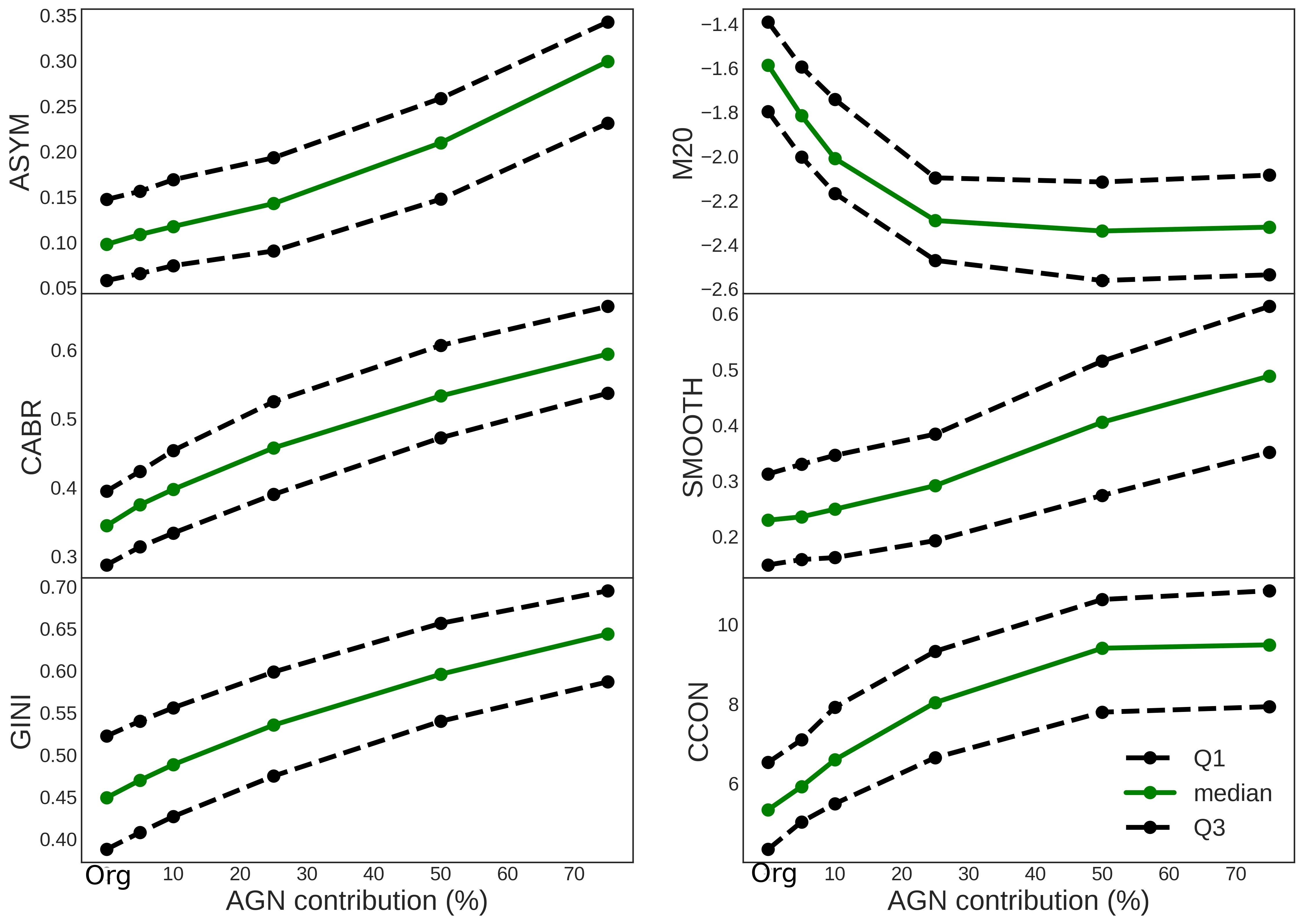}
\caption{Same as Fig. \ref{sec:figlocal}, but only for late-spiral galaxies.}
\label{sec:AGNSpL}
\end{center}
\end{figure}

\begin{figure}
\begin{center}
\includegraphics[width=\columnwidth]{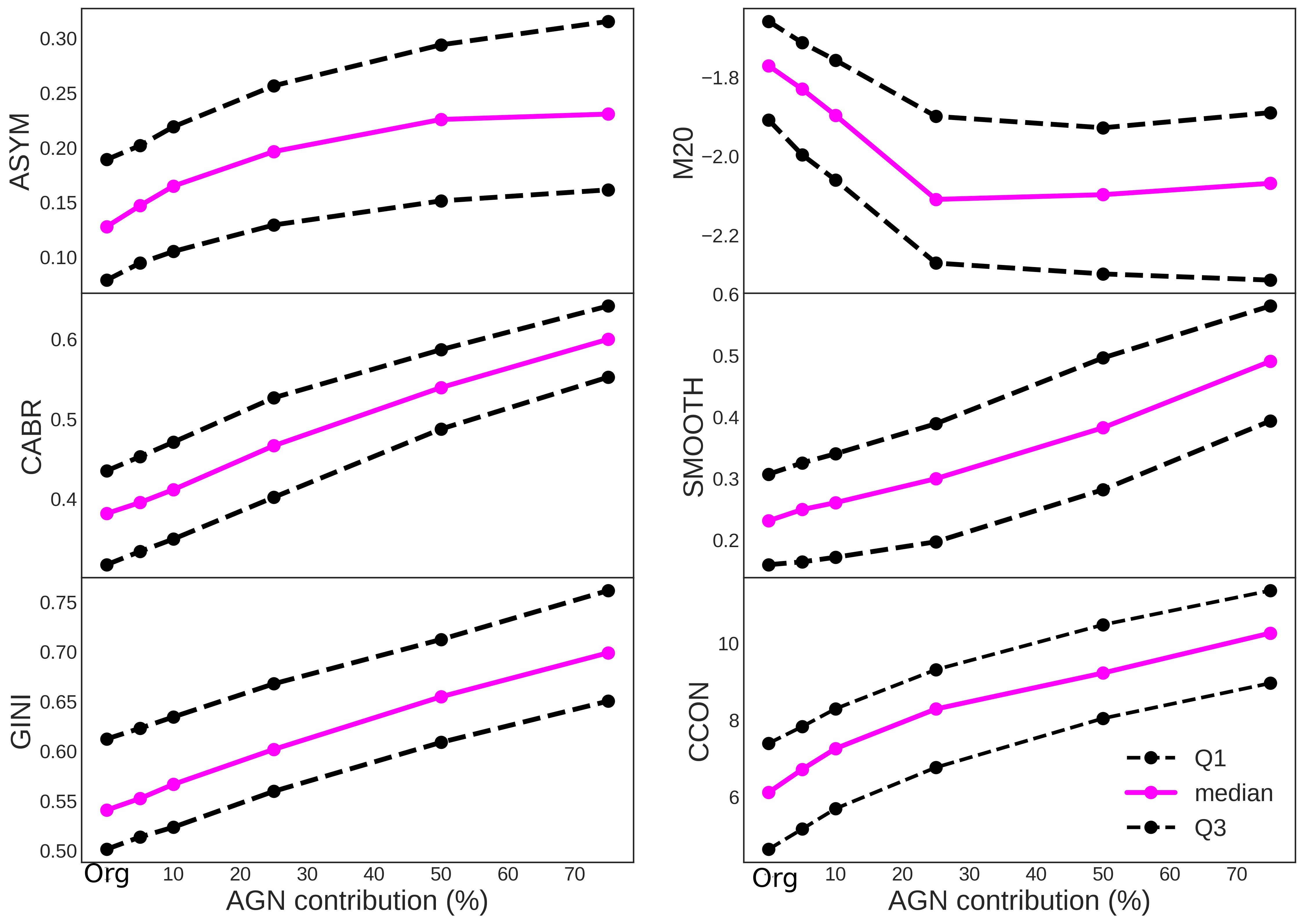}
\caption{Same as Fig. \ref{sec:figlocal}, but only for peculiar galaxies.}
\label{sec:AGNirr}
\end{center}
\end{figure}

Figures~\ref{sec:Ell} to \ref{sec:AGNirr} show how the values of the six
morphological parameters change when including an AGN contribution from 0\% to
75\% of early-type, early-spiral, late-spiral, and peculiar galaxies,
respectively. Table~\ref{table_quantification} provides the
quantification of the change in these parameters for each morphological type
depending on the AGN contribution.

\begin{figure}
\begin{center}
\includegraphics[width=\columnwidth]{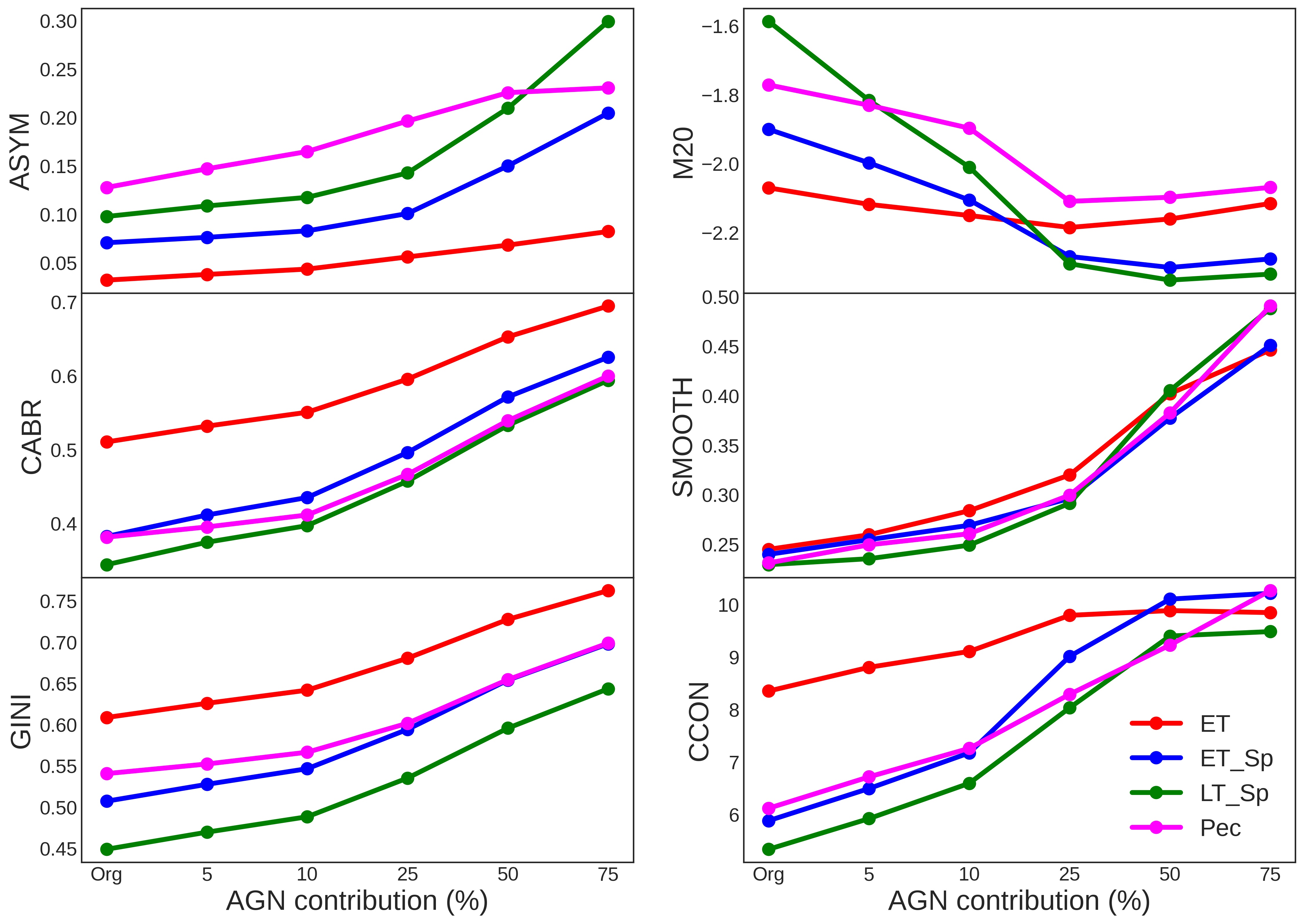}
\caption{Comparison of the effect of 5\%\,-\,75\% of AGN contribution on early-type (in red), early-spiral (in blue), late-spiral (in green), and peculiar (in magneta) galaxies. Median values of each sample are used to represent each distribution, as shown in Figs.~\ref{sec:Ell} to \ref{sec:AGNirr}.}
\label{fig_all_morph_types}
\end{center}
\end{figure}

Table~\ref{table_quantification} shows that the change in all morphological
parameters with the AGN contribution is different for different
morphological types. In general, when dealing with early-type
galaxies, light concentration parameters such as CABR, GINI, M20, and
CCON are less affected by the AGN contribution than in the case of
late-type galaxies (both early- and late-spirals and peculiar), while SMOOTH is more
sensitive to the AGN contribution when measured for early-type galaxies
than for late-types. Similar as in the case of the total sample, ASYM parameter doesn't change much with AGN contribution, independently on morphology. Besides ASYM, GINI is the most stable parameter to be
used when dealing with morphologies of active galaxies, independently
on the morphological type. In addition to this, M20, CABR, and CCON are
also quite stable for active early-type galaxies for AGN
contributions up to 25\%. In general, the morphological parameters
of late-type galaxies, in particular late-spirals and peculiar,
are more affected by the AGN, especially for AGN contribution
of 25\% and above.\\
\indent Finally, to show the behaviour of the 
different morphological types in relation to AGN contribution, all median values are plotted in
Fig.~\ref{fig_all_morph_types}. 
It can be seen the distribution of early-type, early-spiral,
late-spiral, and peculiar galaxies showed previously in
Figs.~\ref{sec:Ell} to \ref{sec:AGNirr}. This figure can help us to
put constraints on each of the parameters regarding their use in
morphological classification of active galaxies, that will be
discussed in more detail in Sec.~\ref{sec5}.

\subsection{Effect of the AGN contribution on the morphological diagrams}\label{sec43}
\begin{figure*}
\begin{center}
{\includegraphics[height= 2.4in, width=3.4in]{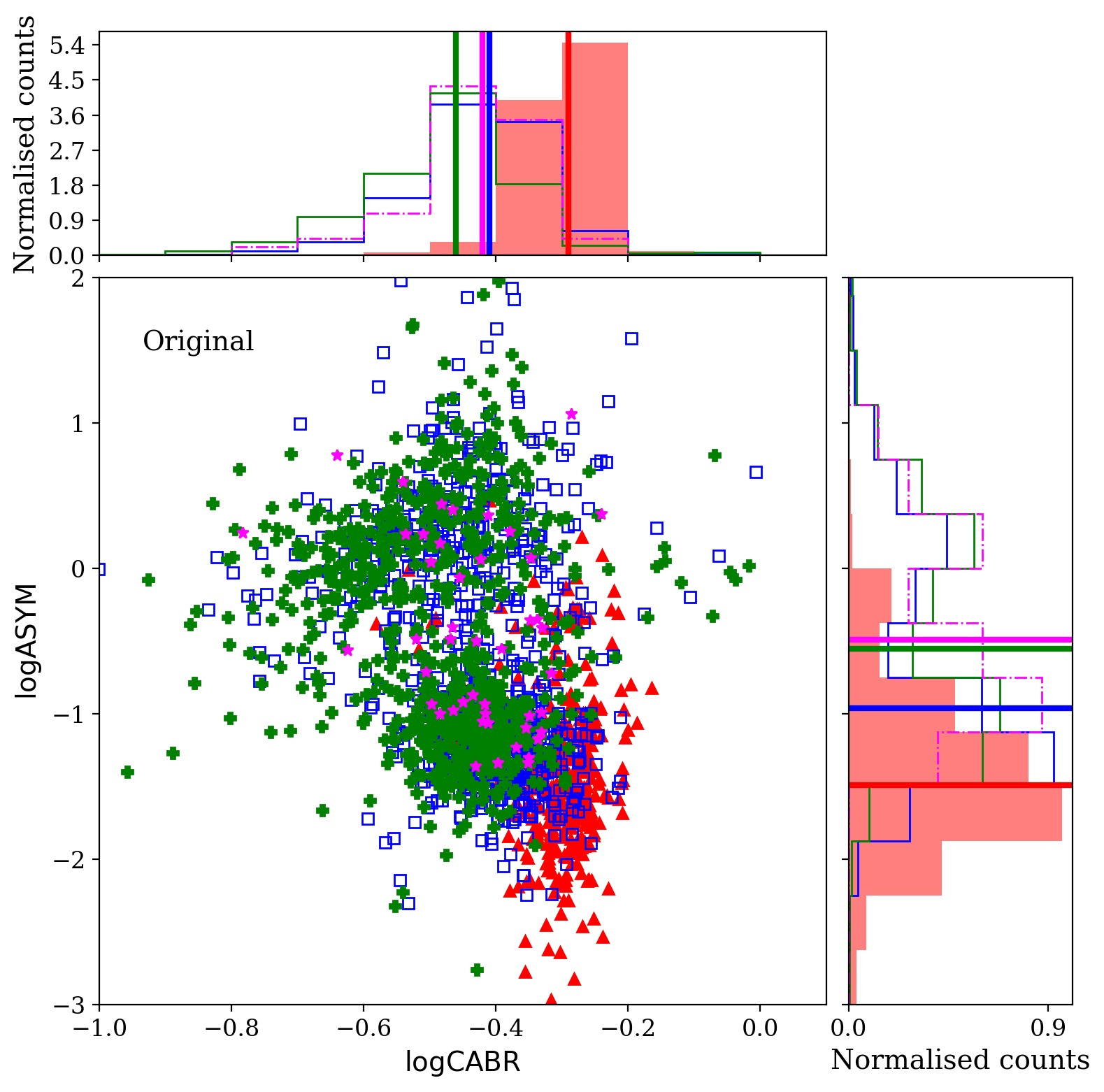}}
{\includegraphics[height= 2.4in, width=3.4in]{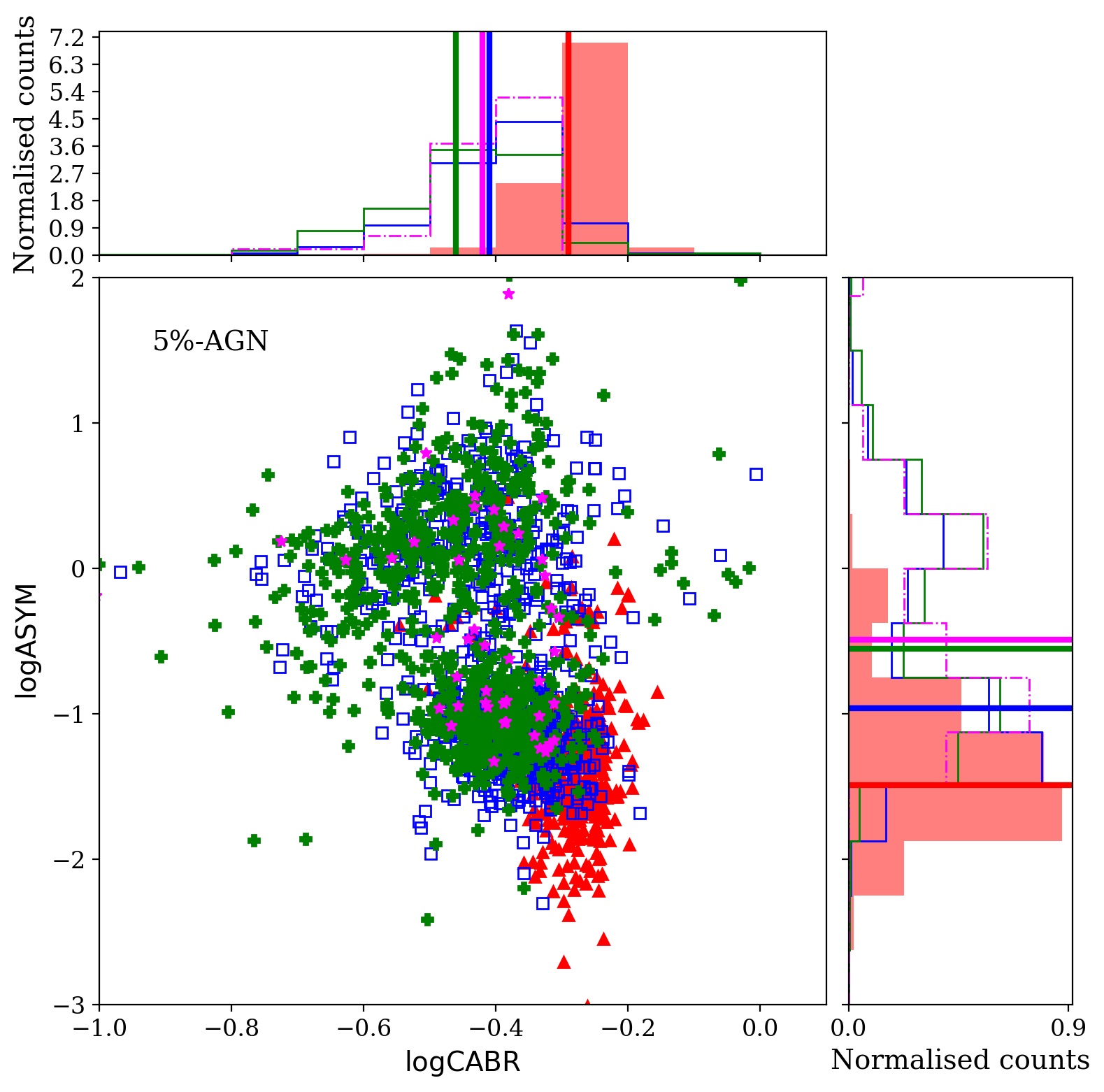}}
{\includegraphics[height= 2.4in, width=3.4in]{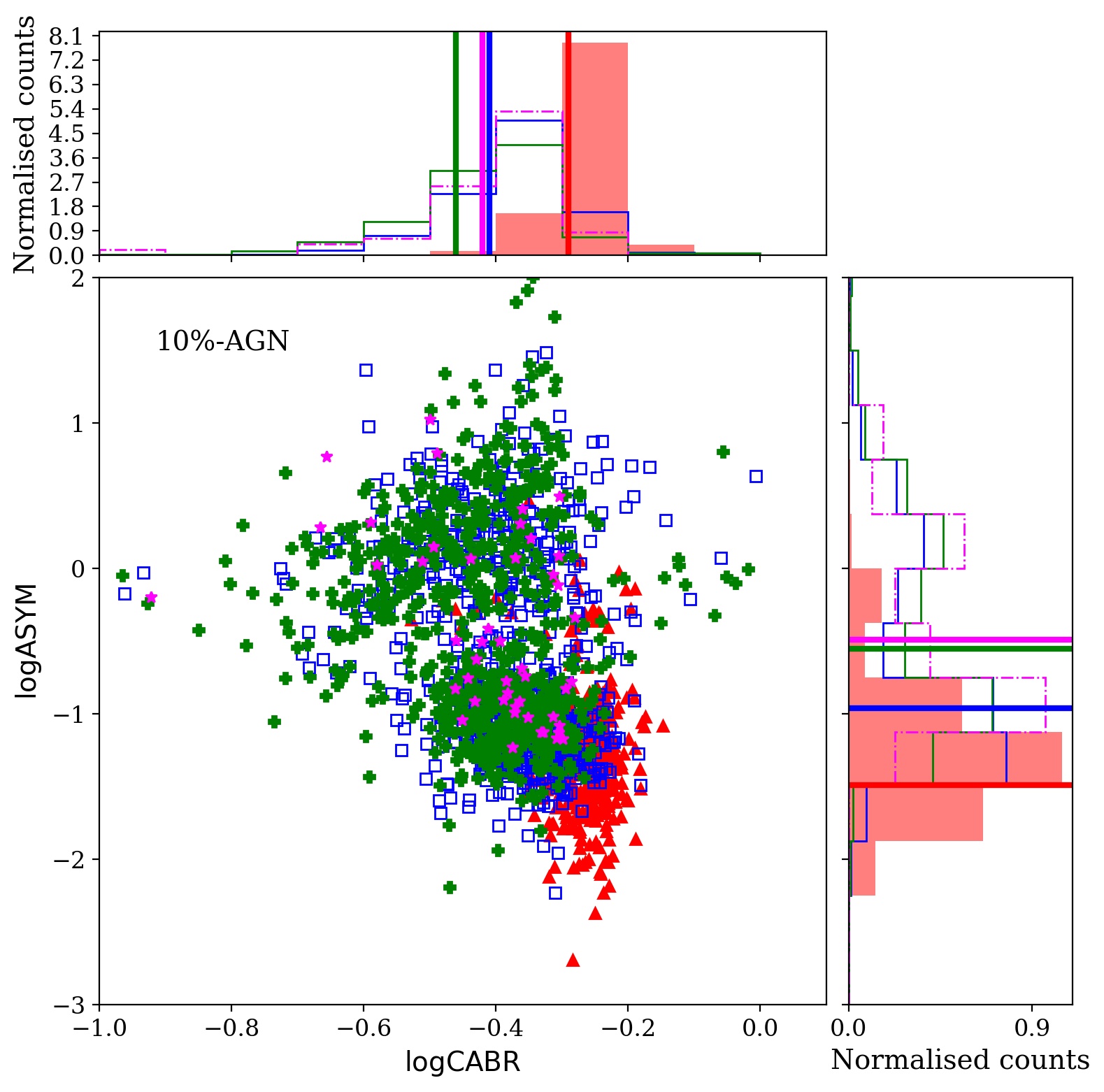}}
{\includegraphics[height= 2.4in, width=3.4in]{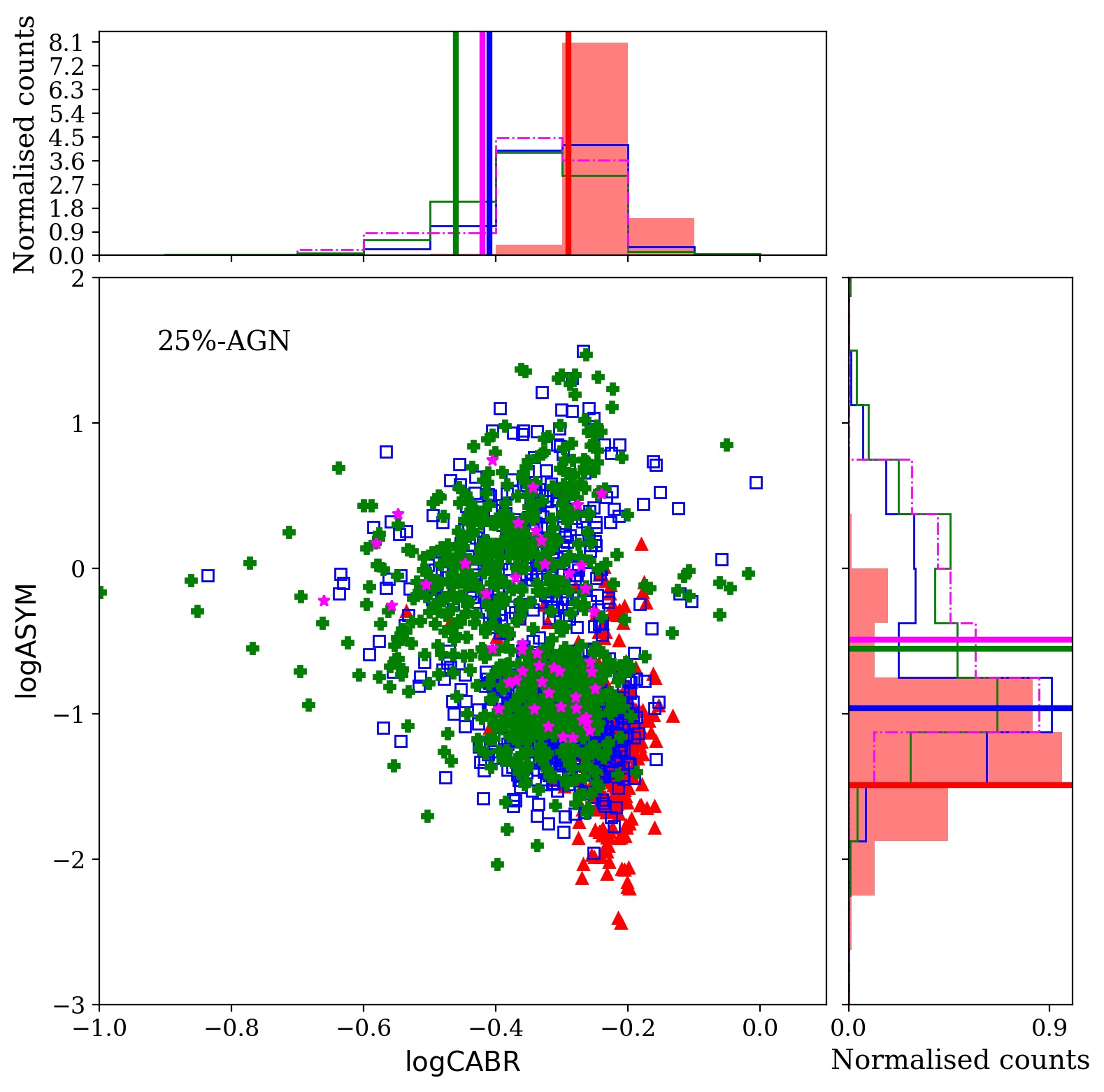}}
{\includegraphics[height= 2.4in, width=3.4in]{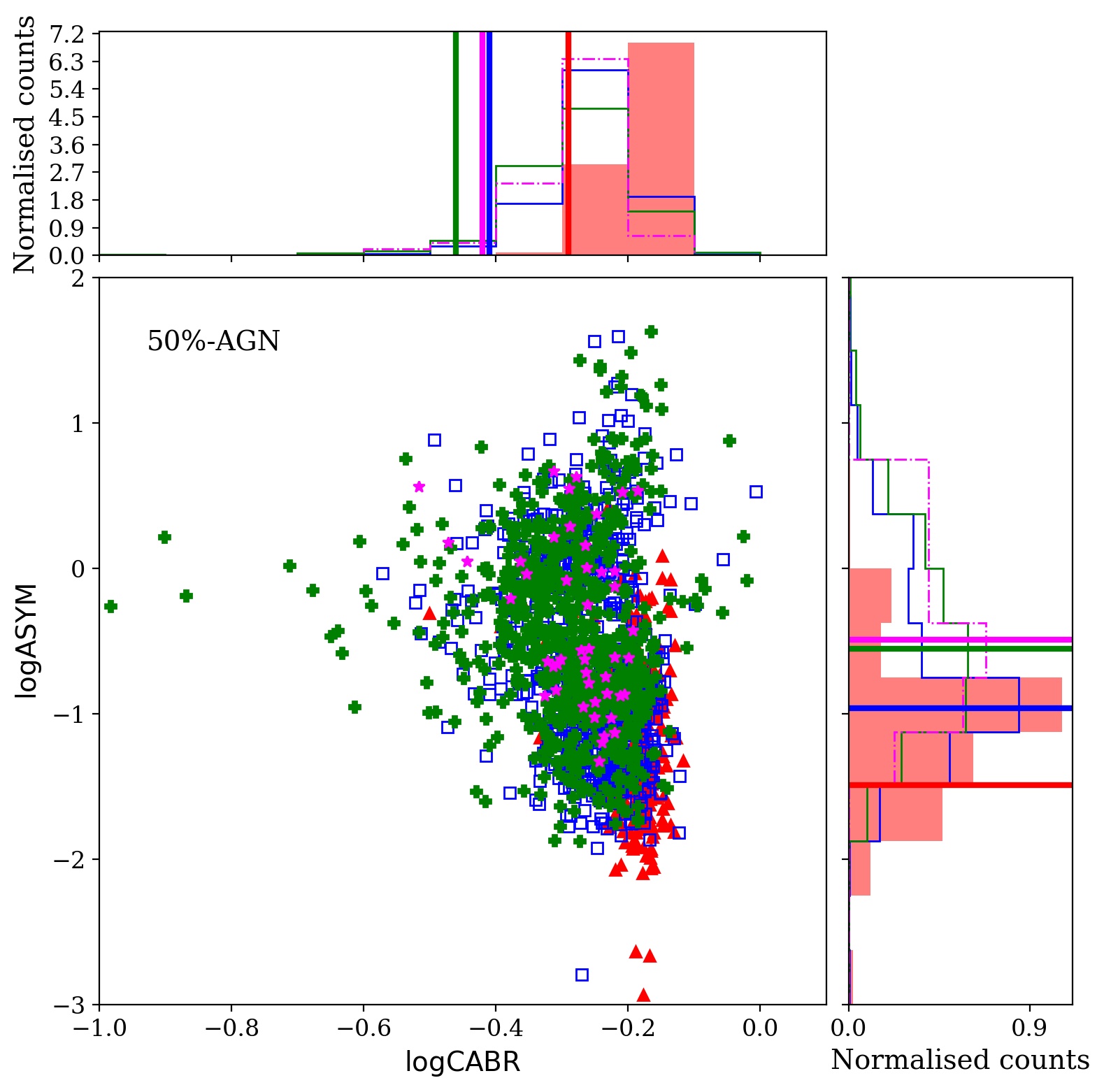}}
{\includegraphics[height= 2.4in, width=3.4in]{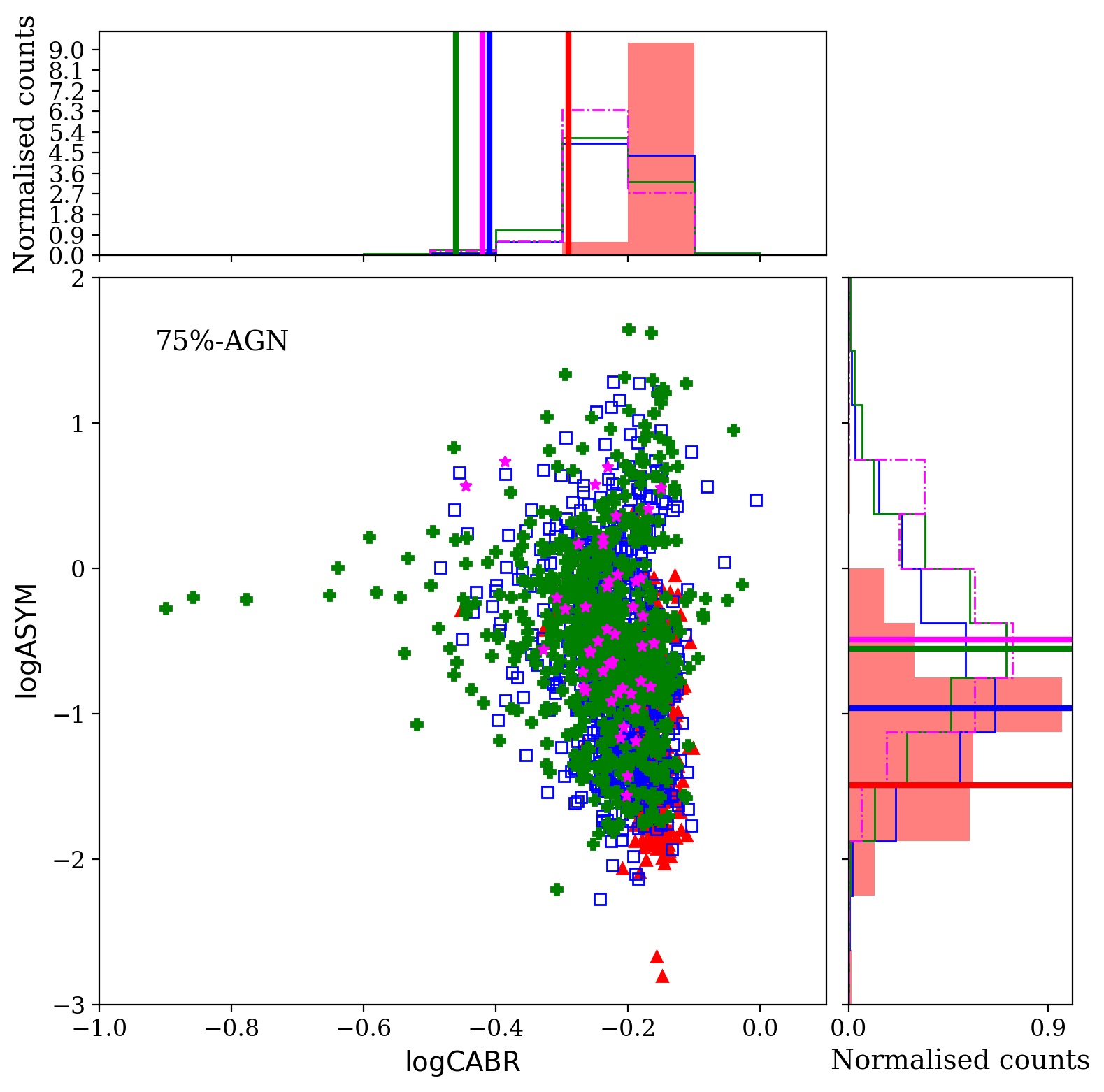}}
\caption{Relation between ASYM and CABR morphological parameters of the original sample of non-active galaxies (top-left plot), and simulated active galaxies with added 5\% (top-right), 10\% (middle-left), 25\% (middle-right), 50\% (bottom-left), and 75\% (bottom-right) of AGN contribution to the total flux. Top and right histograms show the distribution of CABR and ASYM parameters, respectively, of different morphological types. In all, central plot and histograms we represent the distribution of early-type (red), early-spiral (blue), late-spiral (green), and peculiar (magenta) galaxies. In all histograms, as a reference, the red, blue, green, and magenta vertical lines indicate the median values of morphological parameters of the original sample of early-type, early-spiral, late-spiral, and peculiar galaxies, respectively.}
\label{fig_asym_cabr}
\end{center}
\end{figure*}

\begin{figure*}
\begin{center}
{\includegraphics[height= 2.4in, width=3.4in]{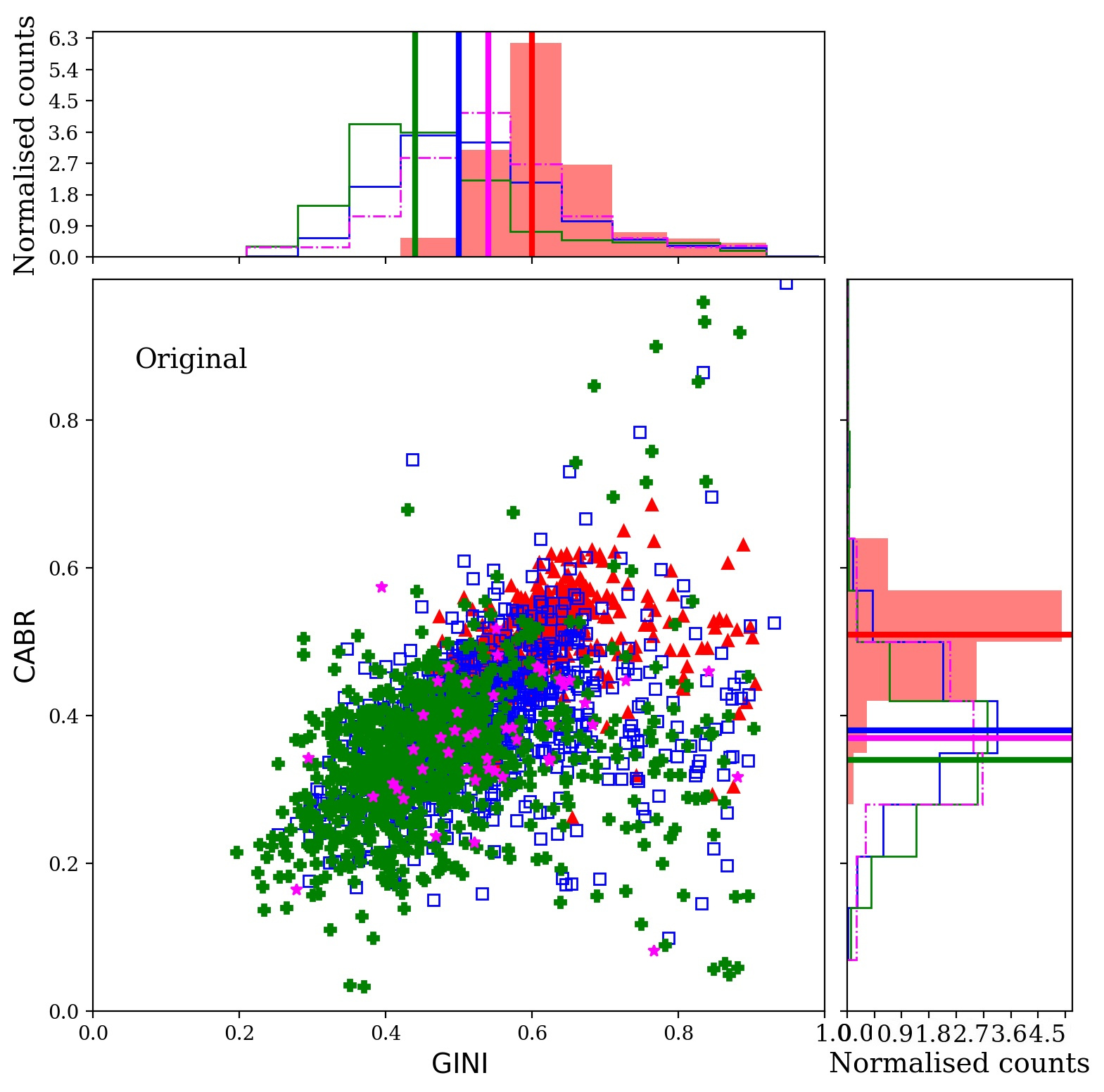}}
{\includegraphics[height= 2.4in, width=3.4in]{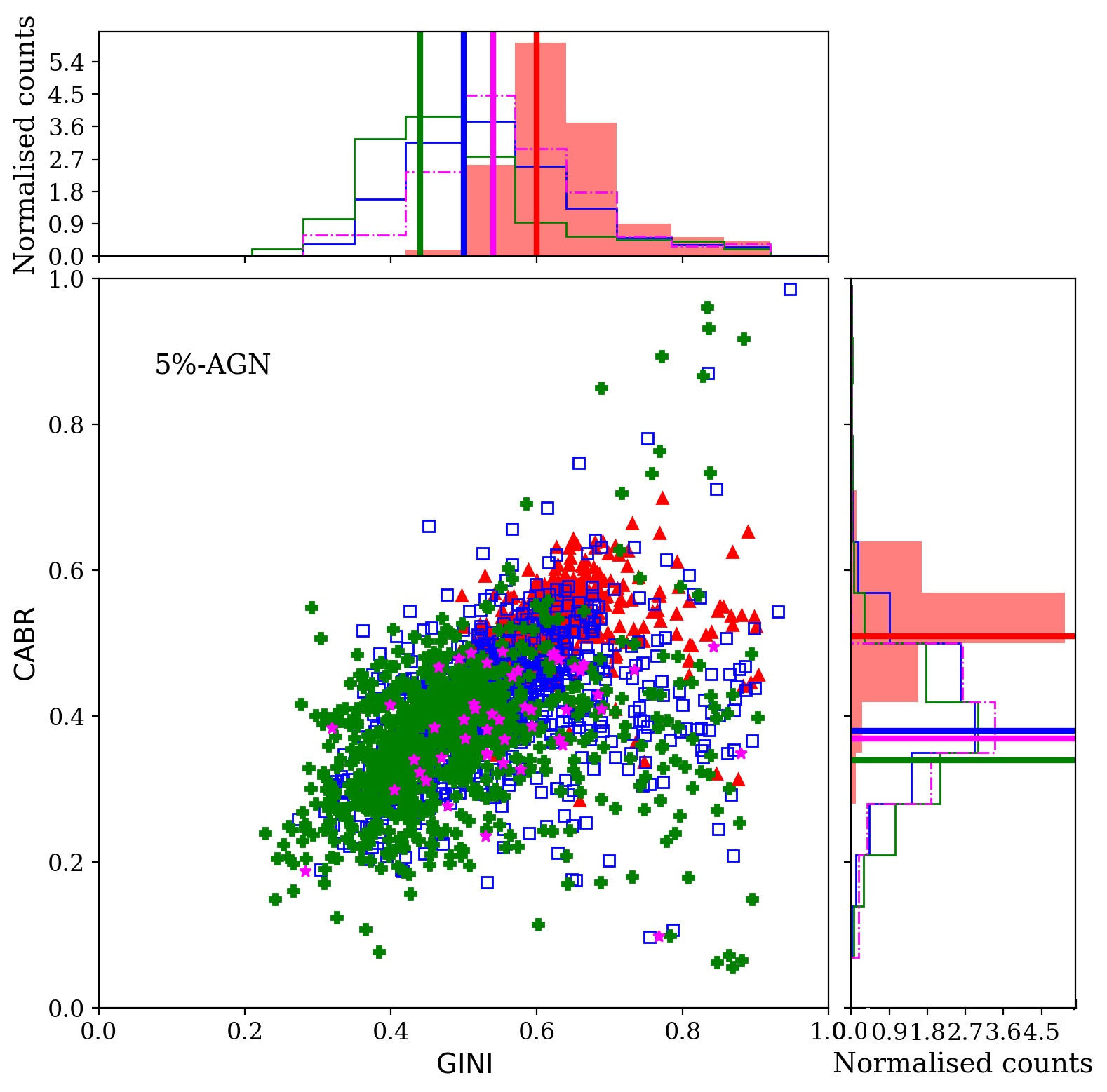}}
{\includegraphics[height= 2.4in, width=3.4in]{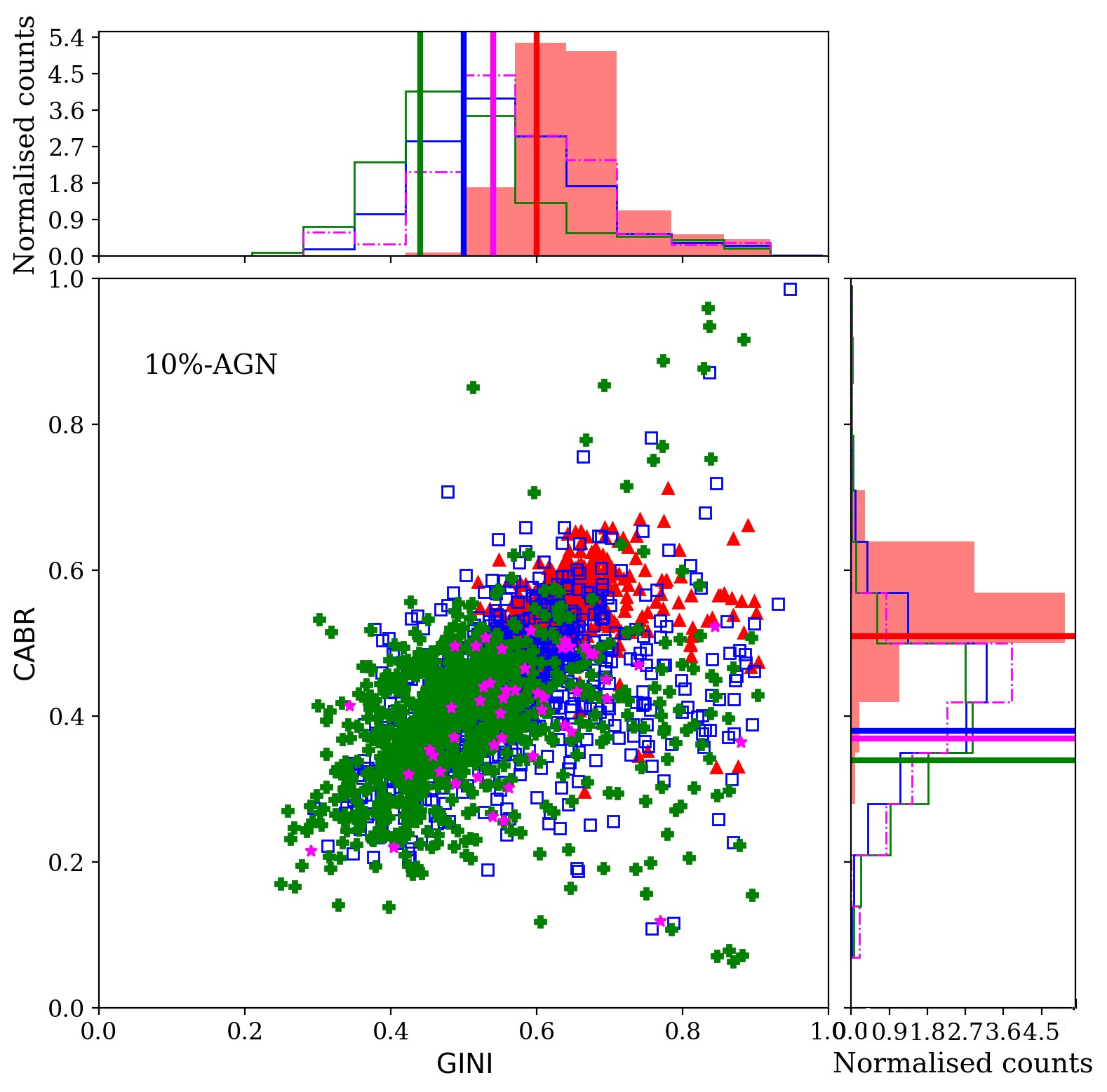}}
{\includegraphics[height= 2.4in, width=3.4in]{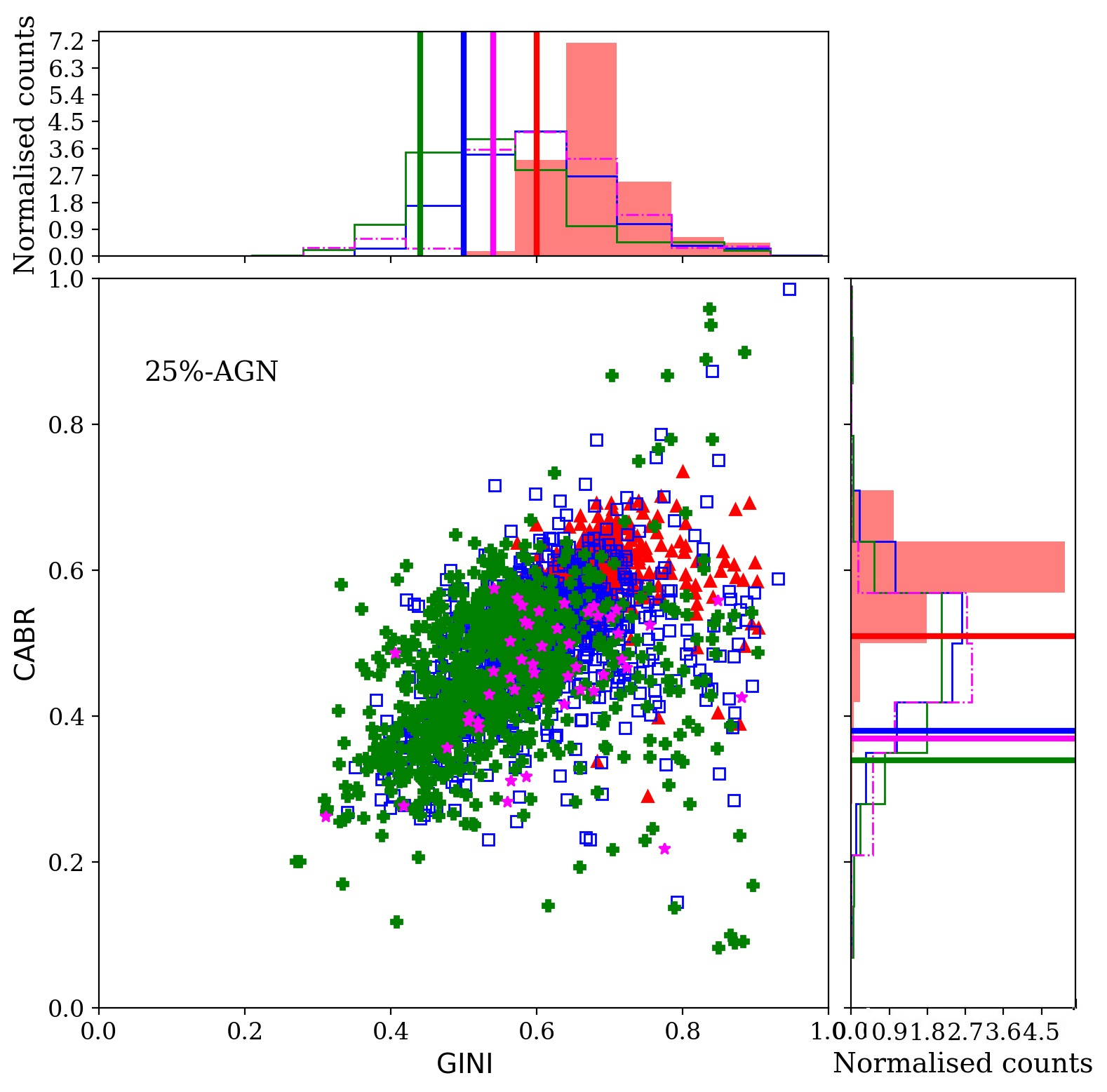}}
{\includegraphics[height= 2.4in, width=3.4in]{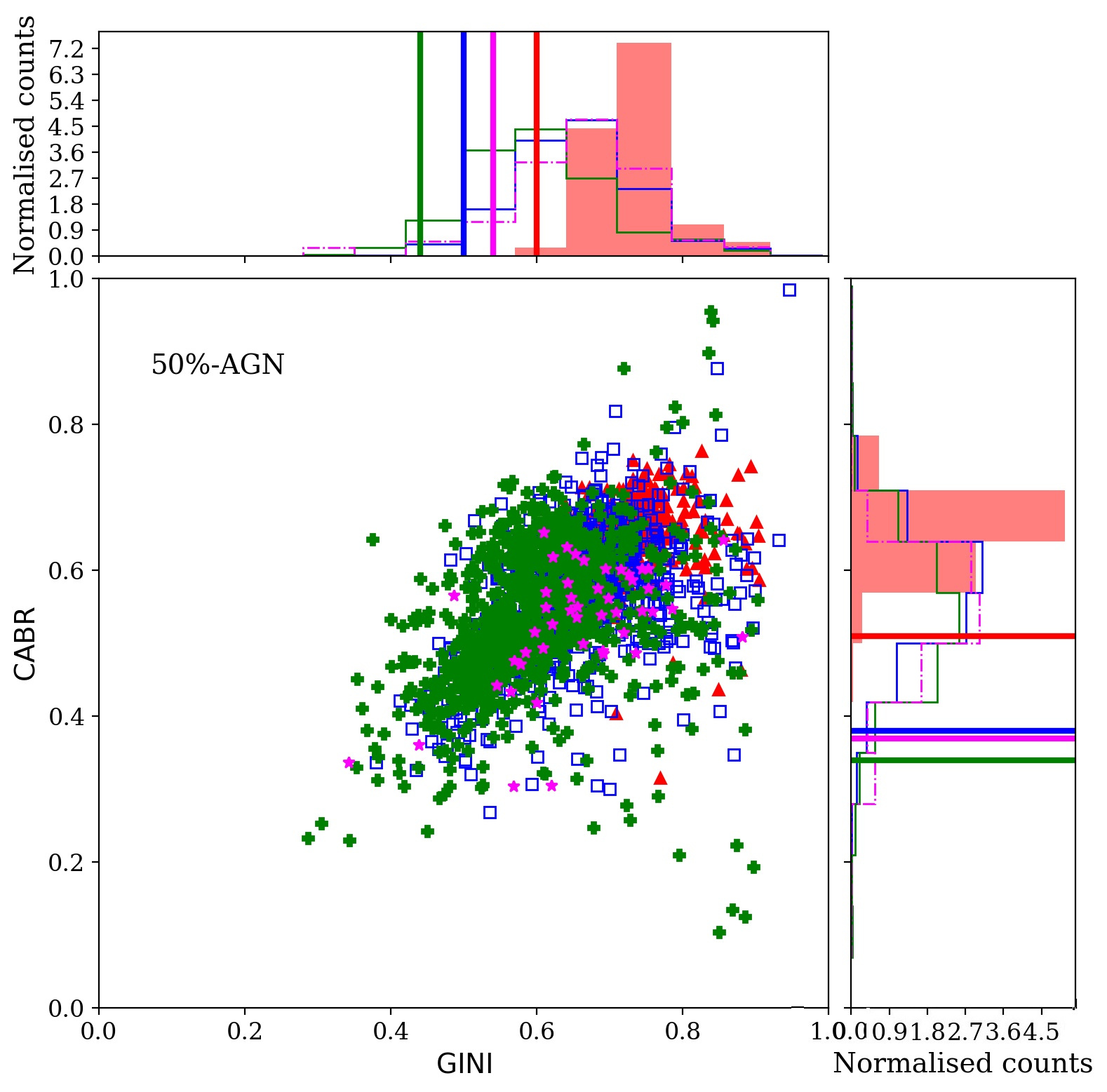}}
{\includegraphics[height= 2.4in, width=3.4in]{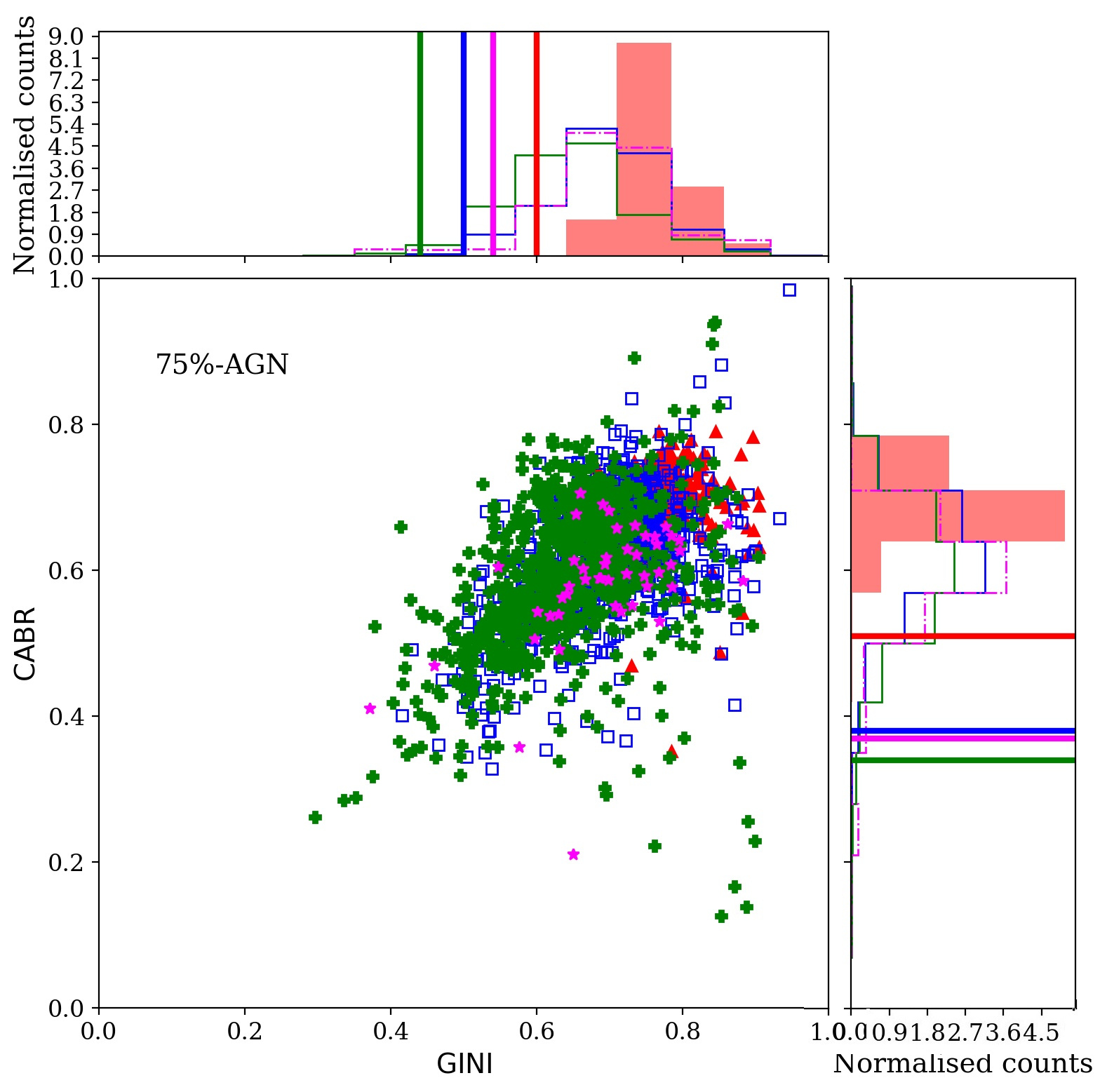}}
\caption{Same as in Fig.~\ref{fig_asym_cabr}, but showing the relation between CABR and GINI parameters.}
\label{fig_gini_cabr}
\end{center}
\end{figure*}

\begin{figure*}
\begin{center}
{\includegraphics[height= 2.35in, width=3.4in]{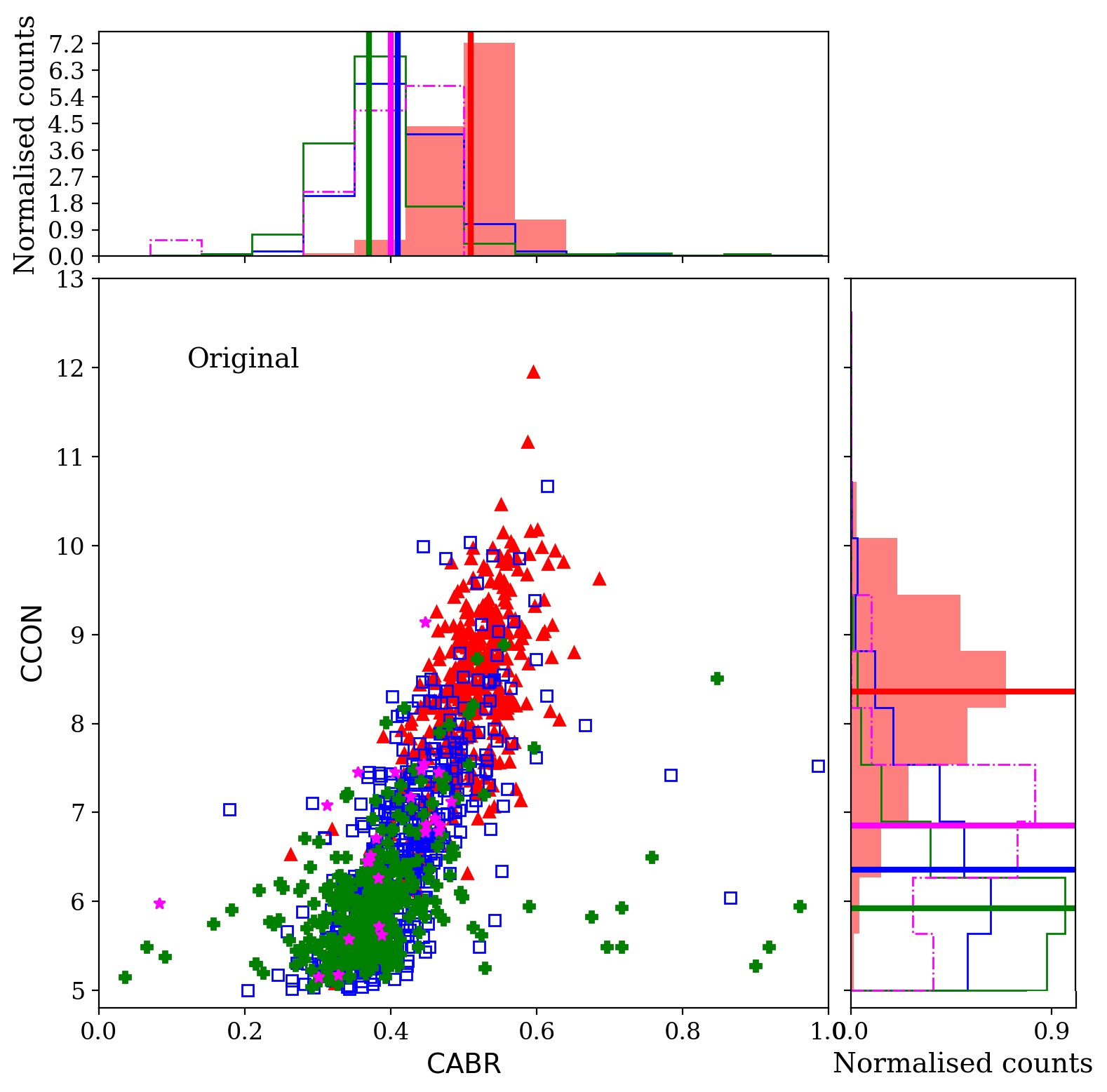}}
{\includegraphics[height= 2.35in, width=3.4in]{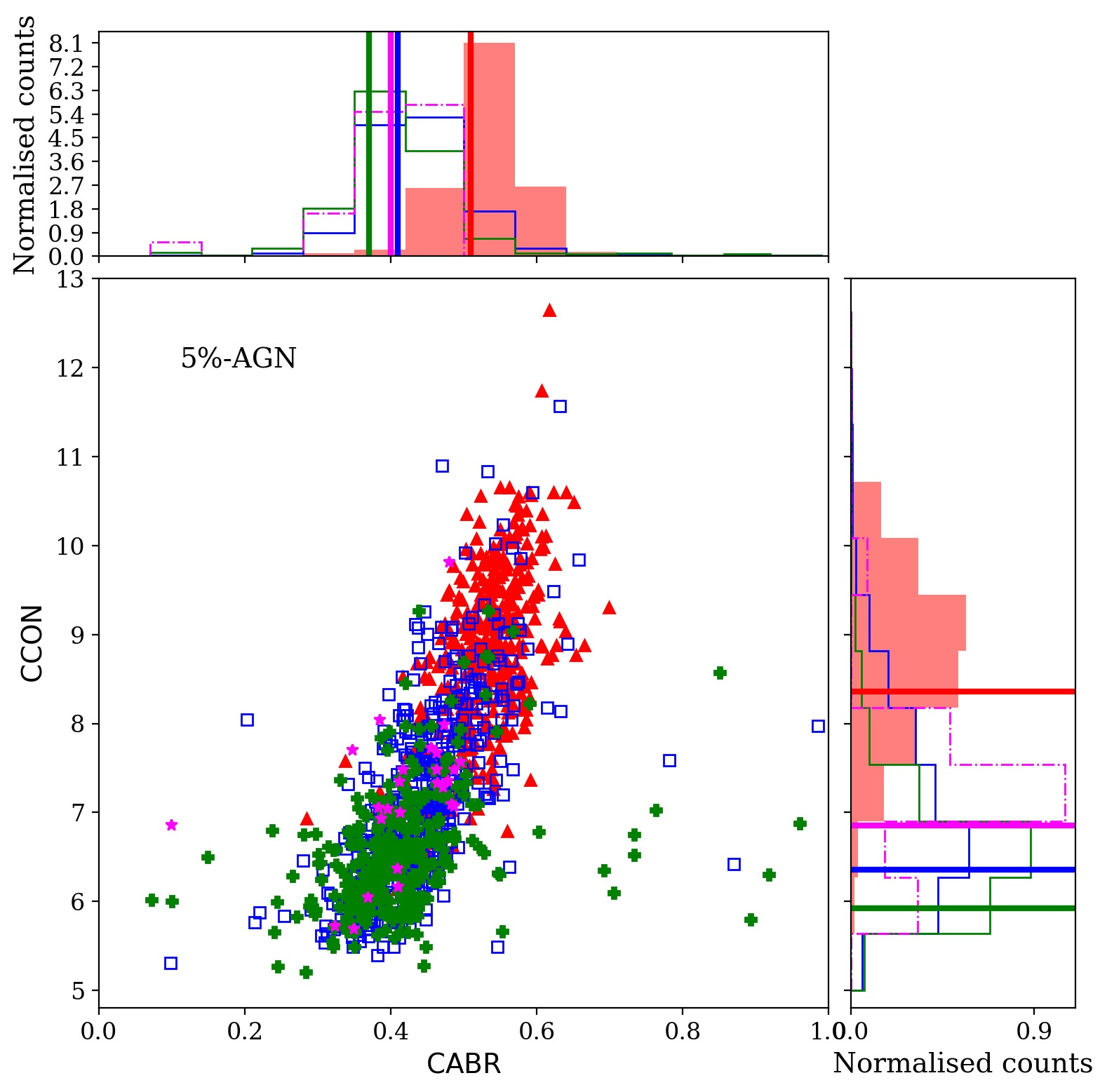}}
{\includegraphics[height= 2.35in, width=3.4in]{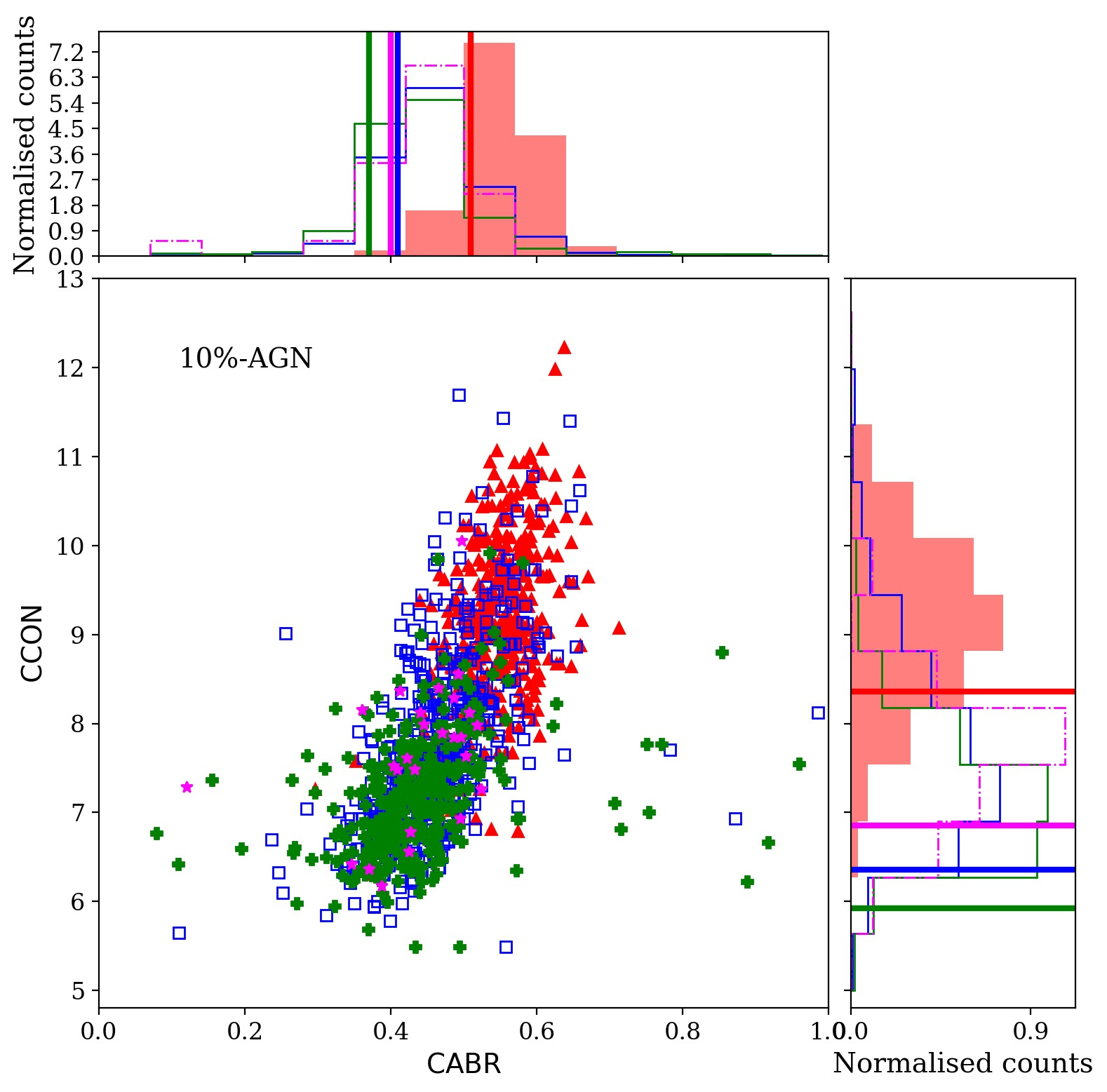}}
{\includegraphics[height= 2.35in, width=3.4in]{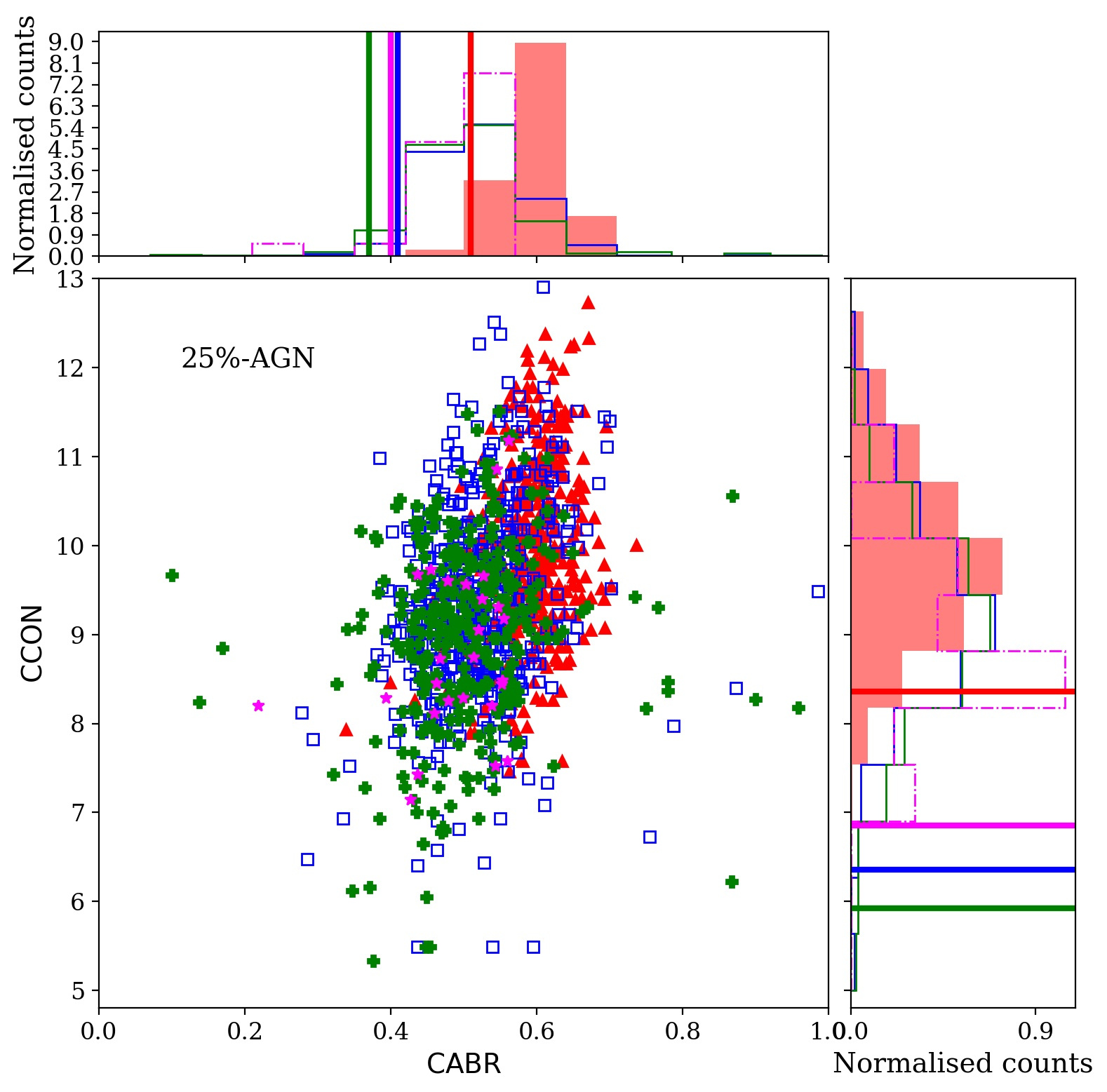}}
{\includegraphics[height= 2.35in, width=3.4in]{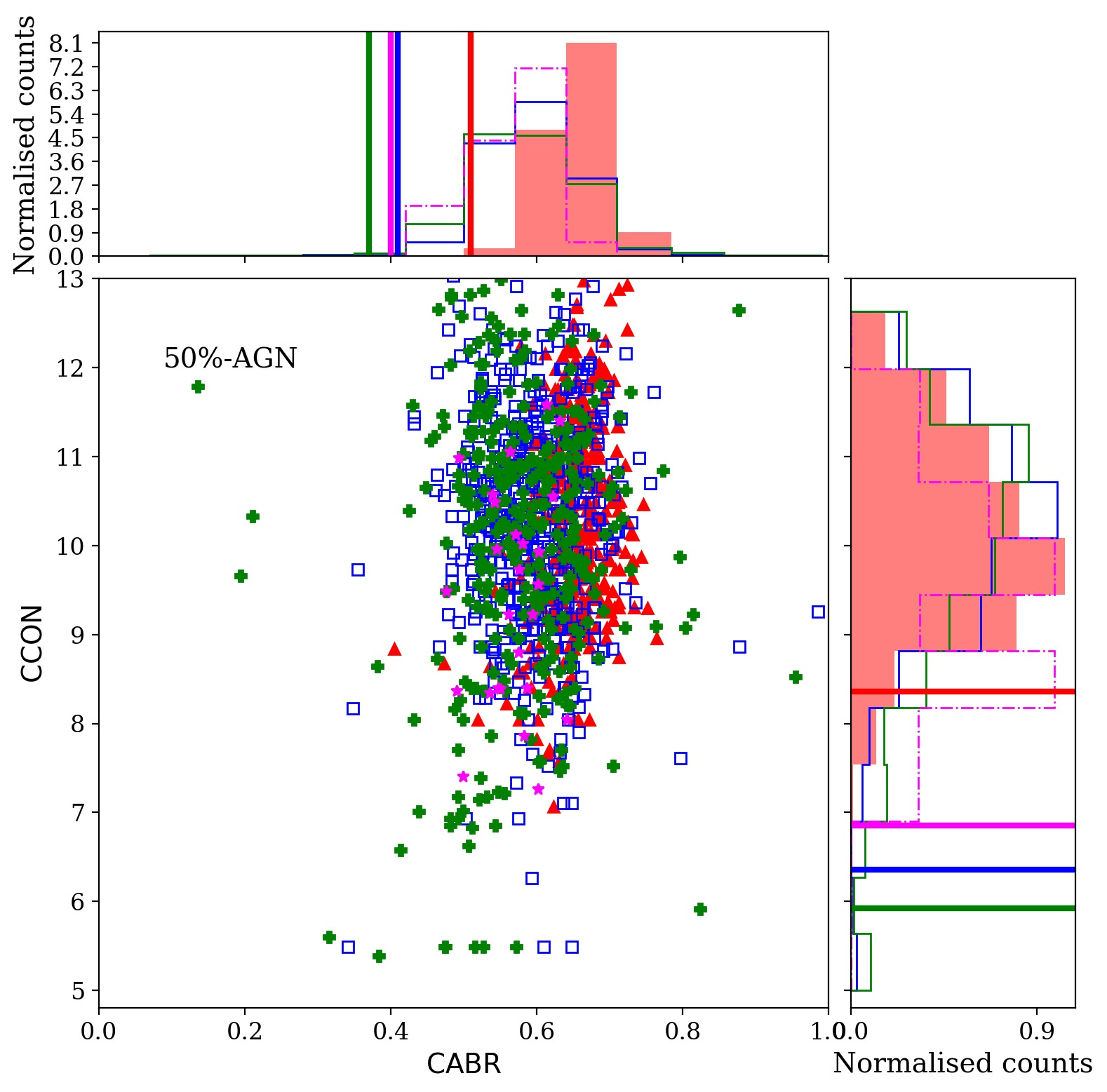}}
{\includegraphics[height= 2.35in, width=3.4in]{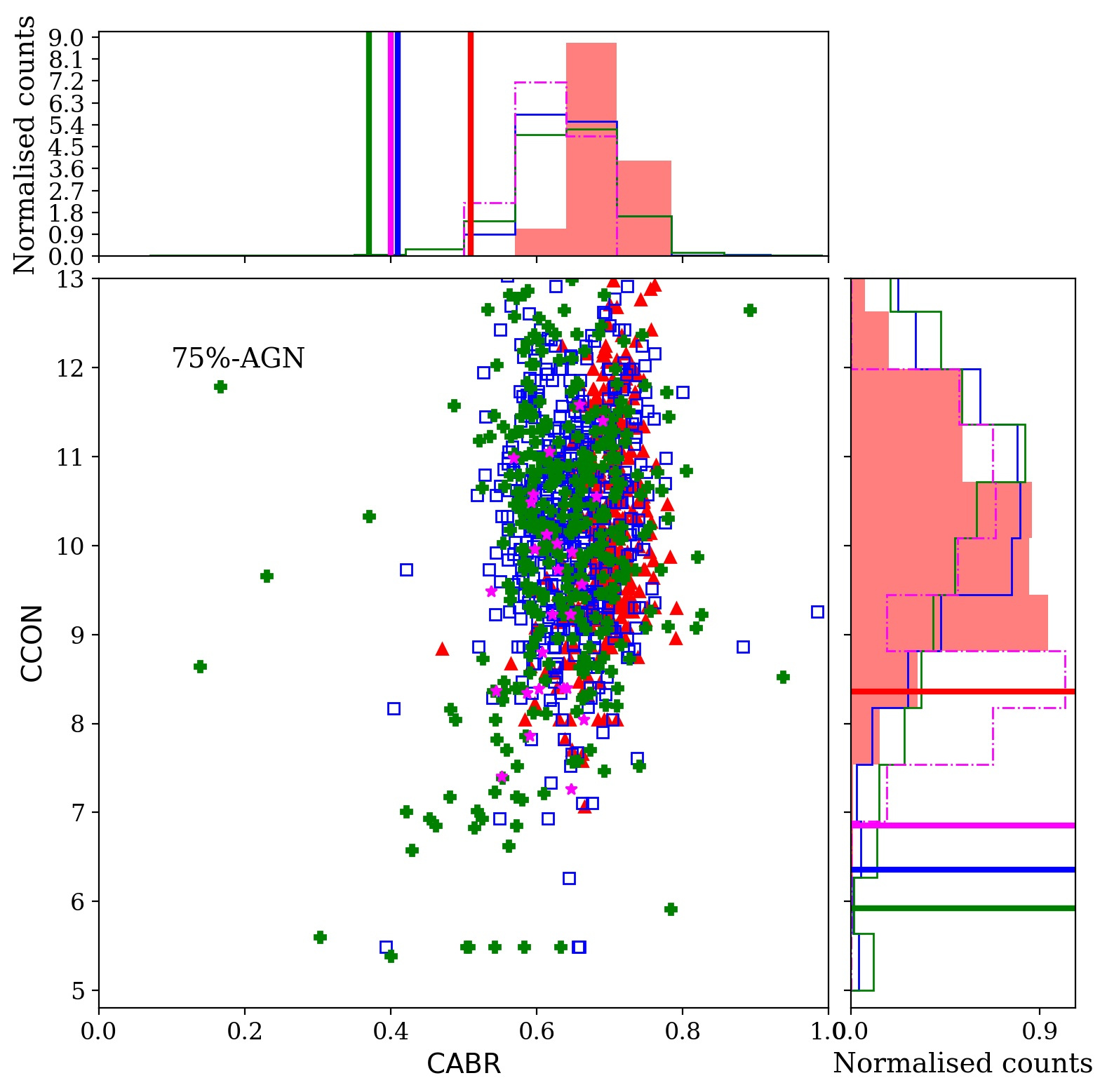}}
\caption{Same as in Fig.~\ref{fig_asym_cabr}, but showing the relation between CCON and CABR parameters.}
\label{fig_cabr_ccon}
\end{center}
\end{figure*}

\begin{figure*}
\begin{center}
{\includegraphics[height= 2.4in, width=3.4in]{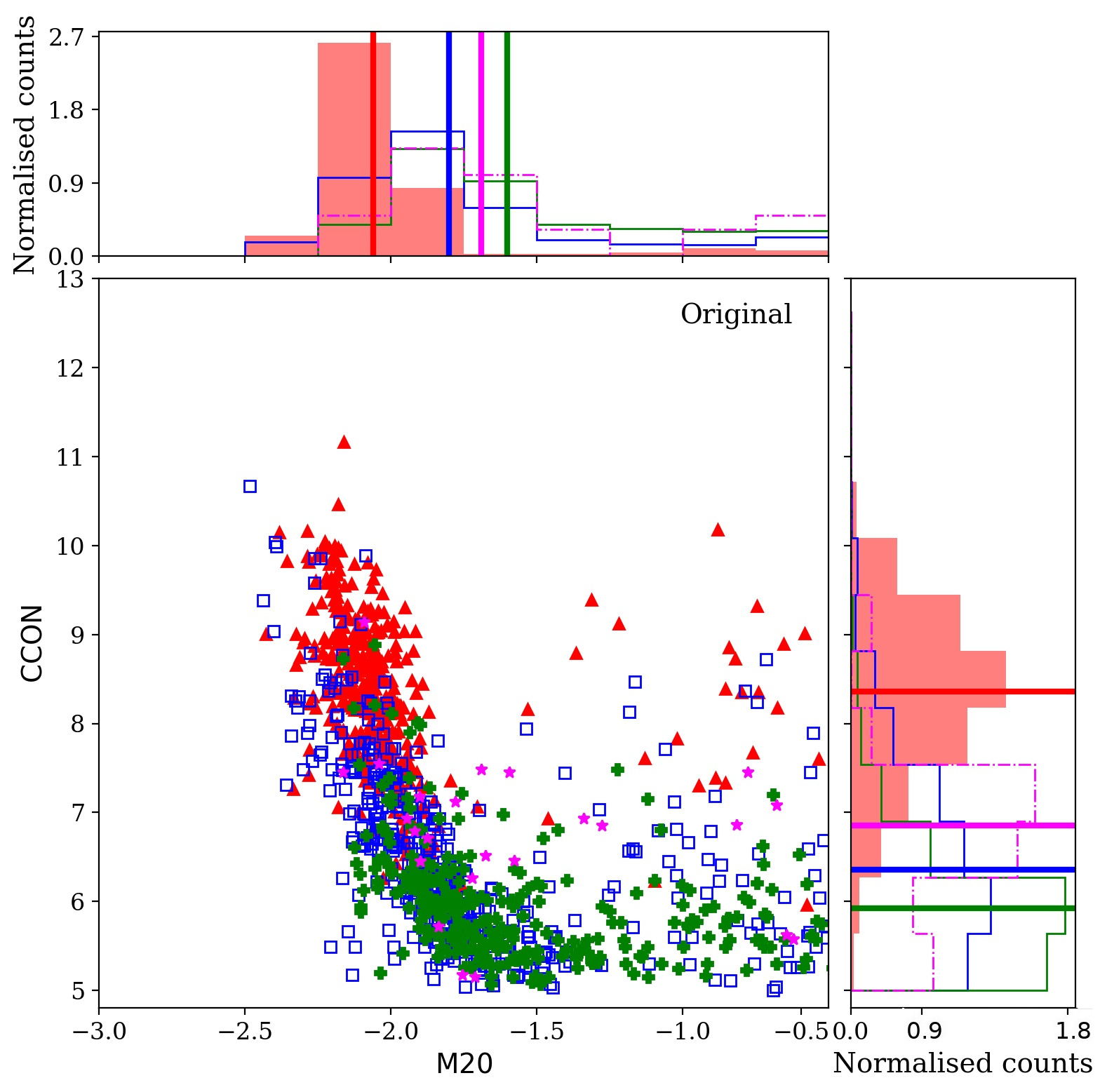}}
{\includegraphics[height= 2.4in, width=3.4in]{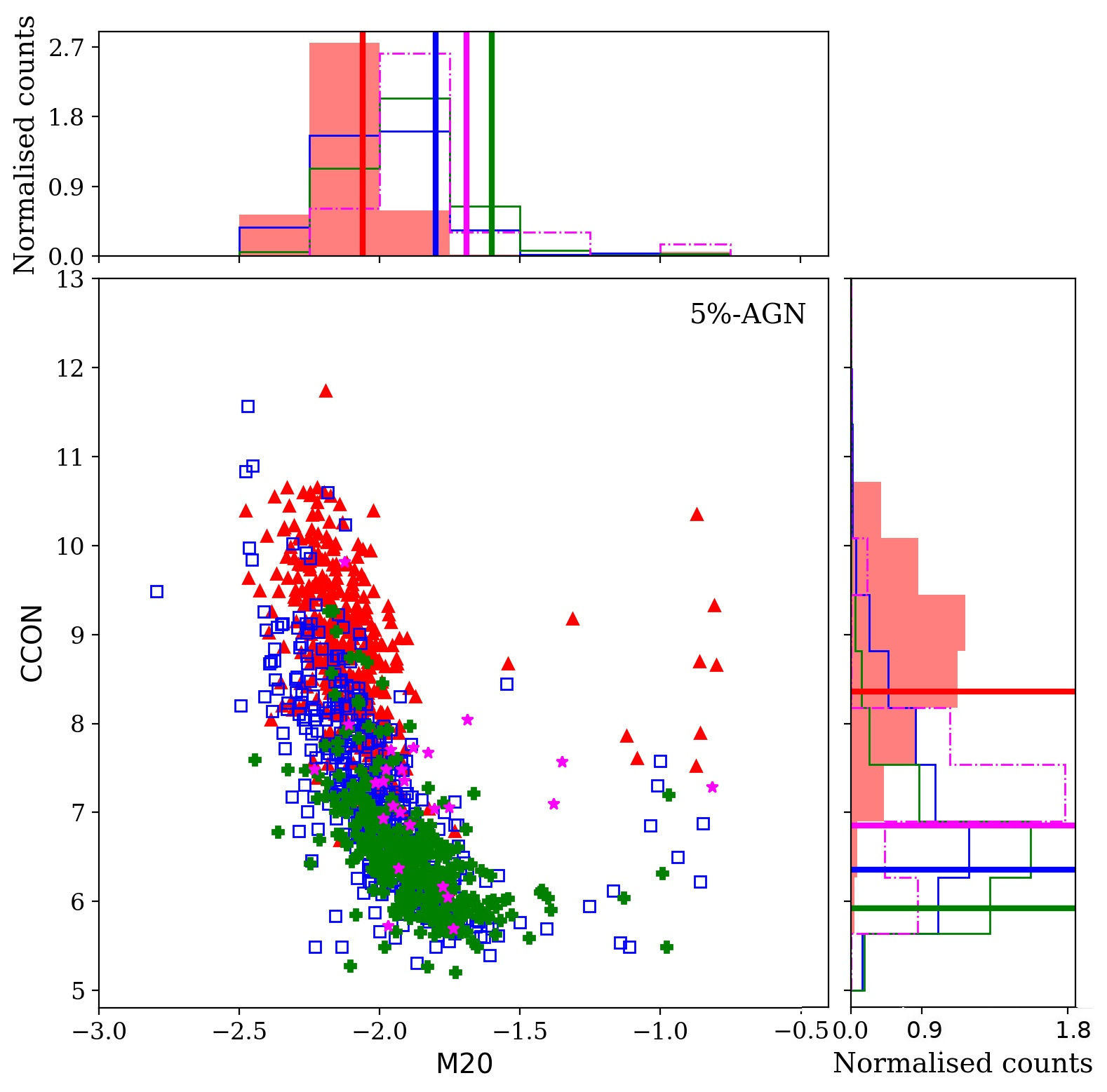}}
{\includegraphics[height= 2.4in, width=3.4in]{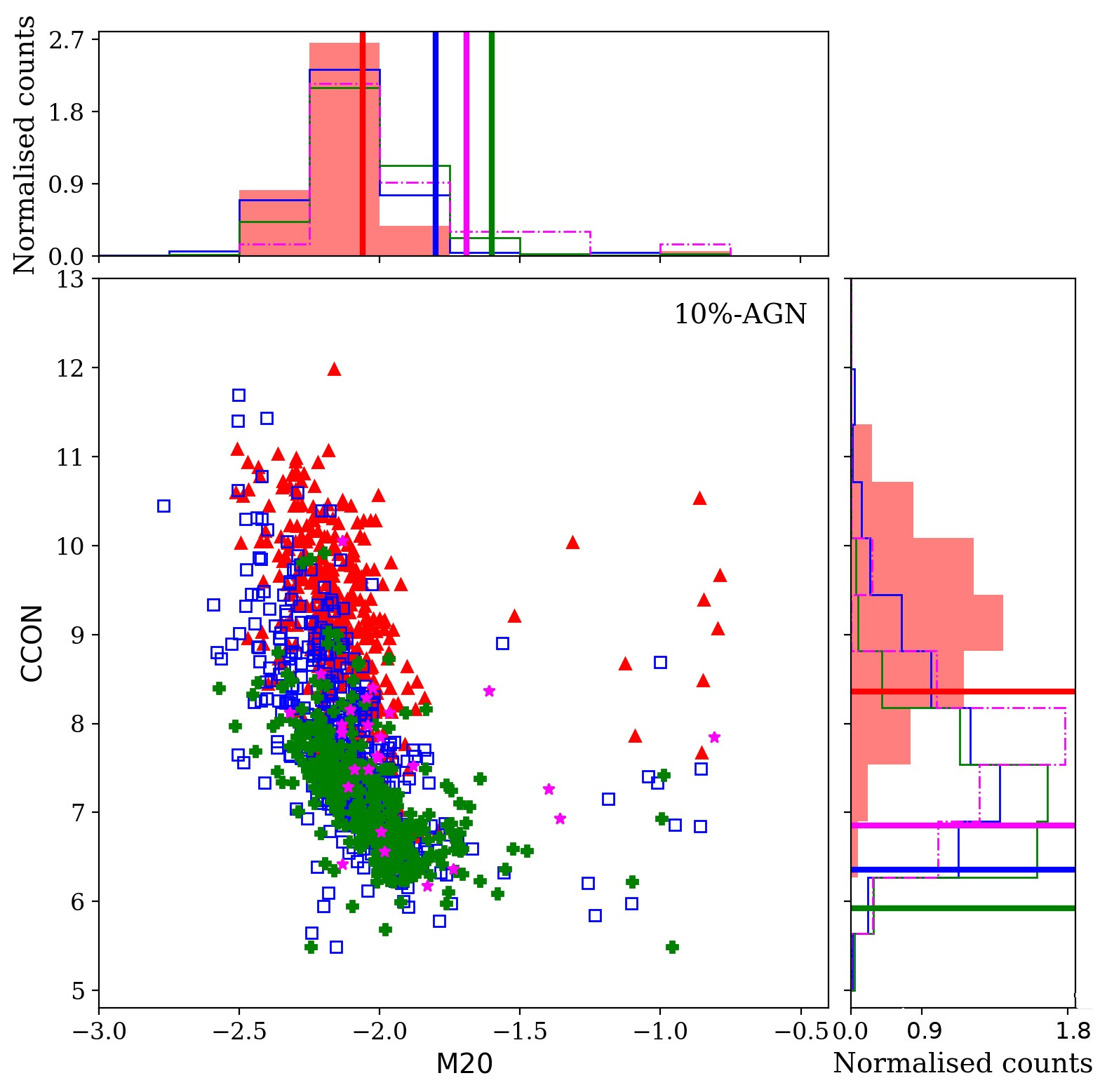}}
{\includegraphics[height= 2.4in, width=3.4in]{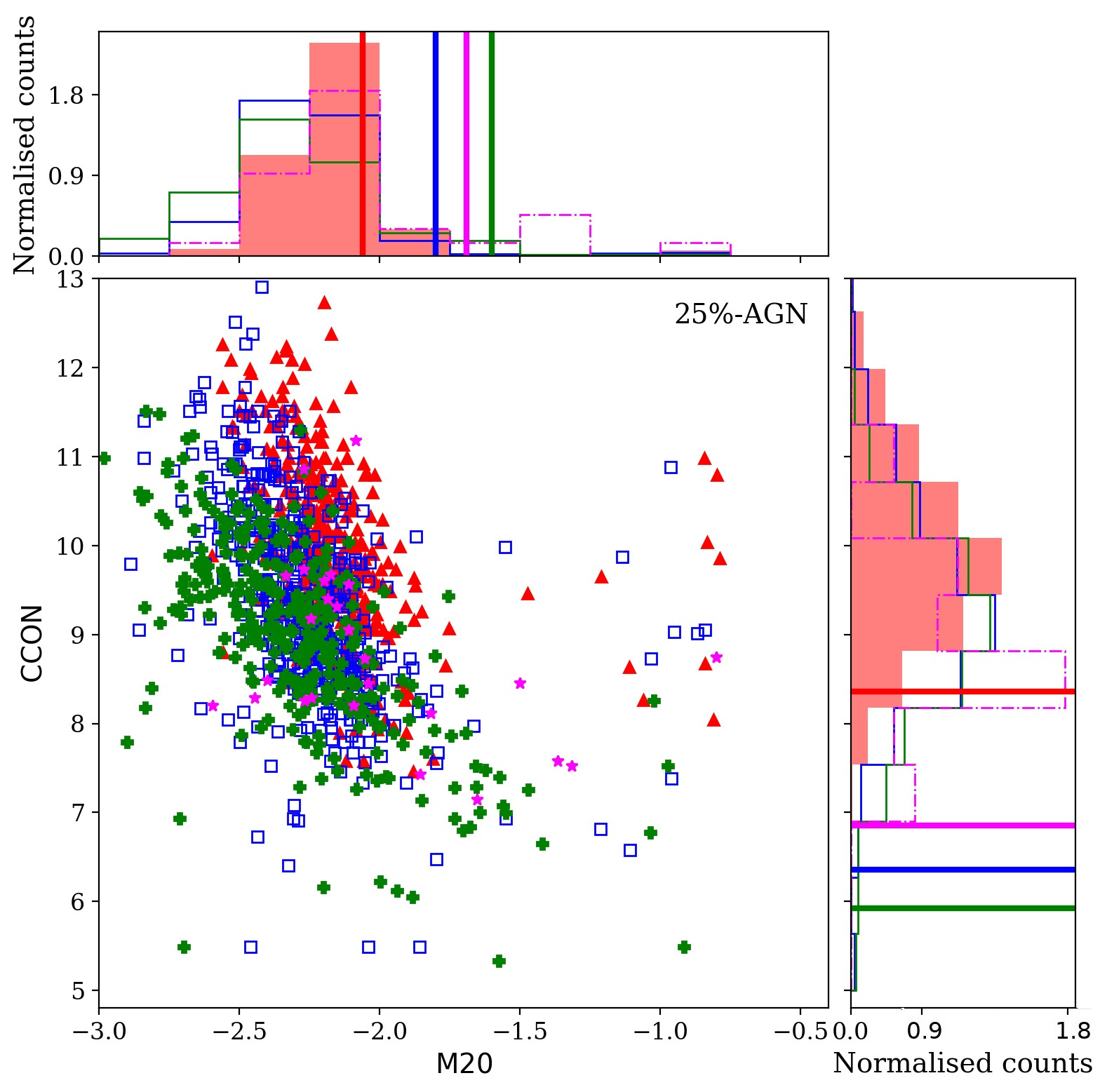}}
{\includegraphics[height= 2.4in, width=3.4in]{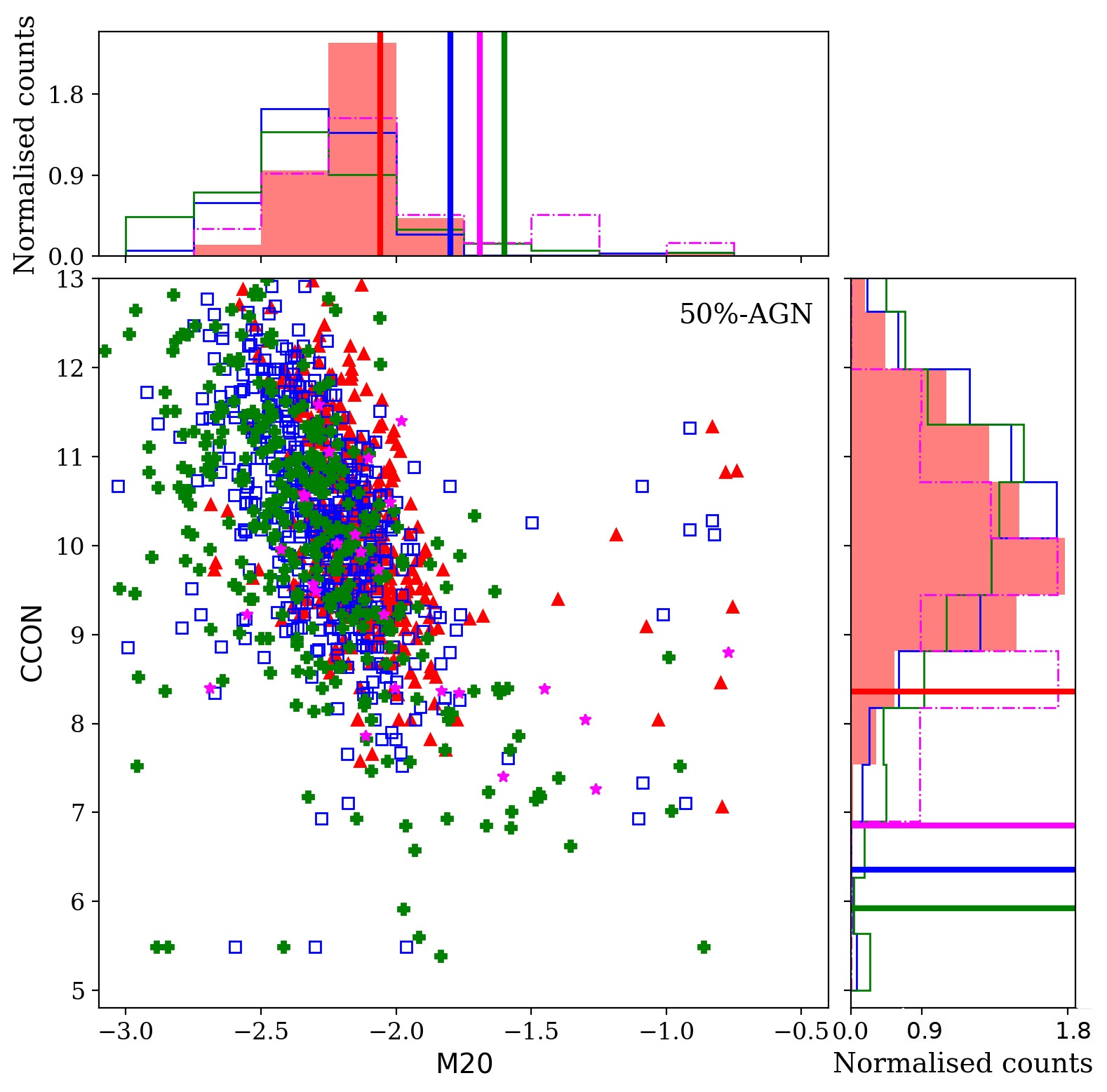}}
{\includegraphics[height= 2.4in, width=3.4in]{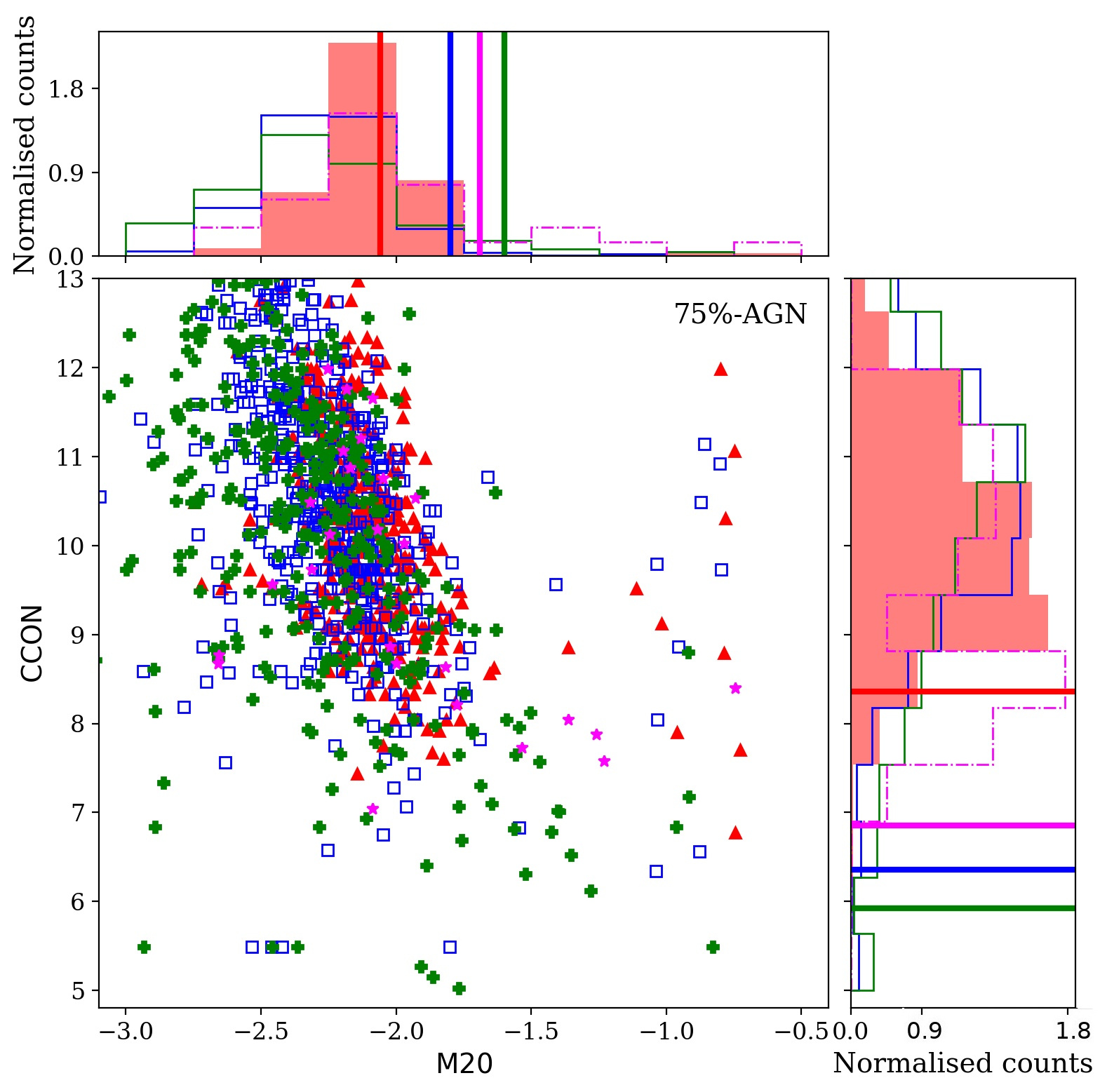}}
\caption{Same as in Fig.~\ref{fig_asym_cabr}, but showing the relation between CCON and $M_{20}$ parameters.}
\label{fig_m20_ccon}
\end{center}
\end{figure*}

In this section, we analyse the effect of the AGN contribution on four of the most commonly
used morphological diagnostic diagrams, that we selected since they
have been broadly used in previous studies to separate between
different morphological types \citep[e.g.,][]{Abraham1994,
  Abraham1996, Abraha2003, Conselice2000, Conselice2003, Lotz2004,
  Yagi2006, Sca2007, Pierce2010, Povic2013, Povic2015, Pintos2016,
  Tars2018, BA2019, Mahoro2019}. Figures~\ref{fig_asym_cabr} to
\ref{fig_m20_ccon} show the following diagrams: ASYM vs. CABR, CABR
vs. GINI, CCON vs. CABR, and CCON vs. M20, respectively. In each
figure, we represent the original local sample of non-active galaxies (top-left plot), and simulated
active galaxies with added 5\% (top-right), 10\% (middle-left), 25\%
(middle-right), 50\% (bottom-left), and 75\% (bottom-right) of AGN
contribution to the total flux. In all plots we represent the location
of early-type (red), early-spiral (blue), late-spiral (green), and peculiar (magenta) galaxies. 
The distribution of each parameter for the four morphological types are also represented as attached histograms on the top and to the right axes of each diagram.
In line with the results obtained in the previous two
sections, in general, it can be observed that for up to 10\% of AGN contribution the
distribution of simulated galaxies does not change significantly in
comparison to the original samples. However, starting from 25\% of AGN
contribution larger contamination can be observed between different
morphological types. Mix between early- and late-types becomes
significantly larger when having strong AGN contribution of 50\% and
75\% to the total galaxy flux, making difficult the separation between these two types as discussed more in Sec.\,\ref{sec5}. For easier comparison with the original sample, in all histograms, as a reference we introduced the vertical lines that correspond to the median values of parameters of different morphological types. Quantification for each parameter is given in Table~\ref{table_quantification}.
		
\section{Results and Discussion}\label{sec5}
As shown in Sec.\,\ref{sec4}., in general, when the light coming from an AGN is $\ge$\,25\% of the total galaxy light, most of the used morphological parameters and therefore the morphological classification of AGN host galaxies can be significantly affected. We found that the classification of early-type galaxies (elliptical and lenticular) hosting an AGN will be more accurate than when dealing with spiral and peculiar hosts. Difficulties in classifying late-type galaxies in comparison to early-types have been already observed in the past, e.g., when studying the effect of survey resolution and depth on the morphological classification of galaxies \citep[e.g,][and references therein]{Povic2015}. \\
\indent In general, when introducing the different AGN contributions, ASYM parameter varies very little. However, taking into account previous studies, this parameter is sensitive to noise \citep{Povic2013} and is measured for a smaller sample of galaxies, and therefore in galaxy classification it shall be used in combination with other parameters. In addition to ASYM, we found GINI to be one of the most stable parameters in the case of both early- and late-types, showing variations $<$\,20\% up to the 25\% of AGN, which still permits the separation between early- and late-types. GINI has been also suggested to trace well sub-structures in galaxies \citep[e.g.,][]{Lisker2006}, mergers and interacting galaxies \citep[e.g.,][]{Lotz2004, Lotz2008a, Lotz2008b}, strong-lensed galaxies \citep[e.g.,][]{Florian2016a, Florian2016b}, and to correlate with stellar mass tracing the overall structure of a galaxy better than any other morphological parameter \citep[e.g.,][]{Zamojski2007}. However, it has been also shown that GINI depends strongly on the signal-to-noise ratio and the choice of the aperture within which it is measured \citep[e.g.,][]{Lisker2008, Povic2015}. In addition to this, we saw that light concentration parameters in general (including CABR, CCON, and M20) are less affected by AGN in early-type than in late-type galaxies, and more stable than SMOOTH parameter which shows higher changes when an AGN is present (although lower when dealing with late-type sources), as shown in Table~\ref{table_quantification}. \\
\indent Previous findings refer to the observed variation in parameters when different AGN contributions are taken into account. However, not necessarily a stronger variation means that a parameter is less suitable for morphological classification. We can observe this in Fig.~\ref{fig_all_morph_types}. ASYM parameter shows to have well separated median values of all four morphological types even up to 50\% of the AGN contribution. CABR and GINI also shown to be efficient for separating early- and late-type (early-spiral, late-spiral, and peculiar) galaxies independently on AGN contribution, where GINI shows better separation between different late-types up to 25\% of AGN. CCON can be also used in classifying early- and late-type AGN hosts up to 25\% of AGN. M20 shows strong mix between morphological types, starting from 10\% of AGN contribution, while SMOOTH parameter shows to be inefficient in classifying active galaxies. These findings with median distributions of parameters can be compared with those in Figs.~\ref{fig_asym_cabr} to \ref{fig_m20_ccon} where the whole samples are represented and similar trends can be observed. Morphological diagrams based on ASYM vs. CABR (Fig.~\ref{fig_asym_cabr}) and CABR vs. GINI (Fig.~\ref{fig_gini_cabr}) parameters remain much more stable on AGN contribution, reaching levels of contamination of early-types by late-types of $\sim$\,30\,-\,35\% for 25\% of AGN contribution, with a slight increment of 5-10\% when dealing with stronger AGN. We can see in both Fig.~\ref{fig_asym_cabr} and \ref{fig_gini_cabr} that even at 50\% and 75\% of AGN contribution to the total optical light, these diagrams still maintain different distributions of early- and late-type galaxies. On the other hand, when using CCON and/or M20 parameters in Figs.~\ref{fig_cabr_ccon} and \ref{fig_m20_ccon} contamination is $>$\,40\% for 25\% of AGN contribution, while for stronger AGN, distributions of early- and late-type galaxies almost totally overlap. Therefore, taking all previous into account, we recommend GINI, CABR, and ASYM to be more suitable to be used in non-parametric methods of morphological classification of AGN host galaxies, even if combined with machine learning. However, we remind once again that for stronger AGN (above 25\% of contribution to the total light), special attention should be taken into account for a reliable classification of AGN hosts at z\,$\sim$\,0, considering the possibility to quantify the AGN contribution to the total optical light of a galaxy and to remove the AGN contribution before going through any detailed morphological classification.\\
\indent To our knowledge, only a few of the previous studies considered the effect of the AGN contribution on the morphological classification of their host galaxies. \cite{Cardamone2010} studied the effect of AGN on the rest-frame colours using a sample of X-ray detected obscured and unobscured AGN in the Extended Chandra Deep Field-South (ECDF-S) at 0.8\,$\le$\,z\,$\le$\,1.2. They found an insignificant effect of the AGN on the galaxy colour in the case of moderate luminosity, obscured AGN, with a more significant effect for very blue, very luminous AGN (sources out of the blue cloud being classified as quasars, see their Fig. 1). A similar study considering the effect of AGN on the host colours has been carried out by \cite{Wang2017} for AGN of moderate X-ray luminosity up to the redshift of z\,$\sim$\,2.5; again significant difference is found in the distribution of U\,-\,V rest-frame colour between active and non-active galaxies. These results go in line with several others showing that in moderate luminosity AGN, the AGN contribution to the total optical light is not significant \cite[e.g.,][]{Silverman2008, Xue2010, Simmons2011}.\\
\cite{Gabor2009} studied the effect of AGN on two morphological parameters, asymmetry, and concentration. They used the same asymmetry parameter as defined in our work, finding an insignificant difference between asymmetries of active and non-active galaxies. However, in their concentration index, defined similarly to CCON, but using $r_{90}$ instead of $r_{80}$, the authors found much larger values of the concentration index when an AGN is present, that is in line with our findings and Fig.~\ref{fig_all_morph_types}. The most similar work to ours is that by \cite{Pierce2010}, where they used a sample of $\sim$\,600 non-active galaxies from the AEGIS survey \citep{Davis2007} at 0.2\,$<$\,z\,$<$\,0.6, and analysed the effect of an AGN on four morphological parameters: Gini, asymmetry, Concelice concentration index, and M20 moment of light. They concluded that AGN contributions above 20\% have an impact on most of the morphology measurements that depend on the morphological characteristics of the original galaxies. In comparison to \cite{Pierce2010}, we provide a more detailed analysis, considering a larger sample of testing non-active galaxies, with precise known visual morphologies, testing five different AGN contributions, and adding two more parameters (CABR and SMOOTH) that were not analysed previously, and that very often are used simultaneously with other parameters in the morphological classification of galaxies \citep[e.g.,][]{Huertas2008, Huertas2009, Povic2013, Pintos2016, BA2019, Mahoro2019} \\
\indent Finally, we recall that the results obtained in this paper refer to the AGN effect on the morphological properties of their host galaxies at z\,$\sim$\,0. When going to higher redshifts and fainter magnitudes, and due to the effect that both the survey depth and resolution have on the morphological classification of galaxies, the study will be more complex, as mentioned above and analysed in detail in \cite{Povic2015}. We plan to proceed with this approach in a forthcoming paper (Getachew et al., in prep.).

\section{Summary and conclusions}\label{sec6}
In this paper, we provide a detailed study regarding the effect of the AGN on the morphological properties of their host galaxies and the corresponding morphological classification. We studied six parameters commonly used in non-parametric methods, often in combination with machine and deep learning, to classify galaxies in terms of their morphology. These parameters are: Abraham concentration index (CABR), Gini coefficient (GINI), Conselice-Bershady concentration index (CCON), $M_{20}$ moment of light (M20), asymmetry index (ASYM), and smoothness (SMOOTH). We used the sample of 8000 local galaxies with visual morphological classification in g-band \citep{Nair2010} to select a sample of 2301 visually classified non-active local galaxies. We simulated their AGN counterparts by adding in their centers an AGN contribution of 5\%, 10\%, 25\%, 50\%, and 75\% of the total galaxy flux. For the AGN construction for each galaxy we used its corresponding PSF, described by the Moffat function. We then measured in a consistent way the six morphological parameters in both the original images and all simulated images. We finally studied how the presence and relative contribution of the AGN may affect each of the morphological parameters in relation to galaxy morphology, and the final classification of active galaxies. Our main findings are the following: 
\begin{itemize}
\item When observing the total sample of galaxies, independently on their morphology, we observe that significant change of $\sim$\,30\% in parameters corresponds to AGN contributions to the total optical flux of 25\% and above, except in ASYM which almost doesn't vary when AGN is added.\\
\item An impact of AGN on different morphological parameters vary depending on galaxy morphology. Light concentration parameters (e.g., CABR, GINI, M20, and CCON) are less affected by AGN in early-type galaxies, while SMOOTH is less affected in late-types. However, in general, spiral and peculiar galaxies will be more affected by AGN contribution than early-type galaxies. \\
\item GINI and ASYM have been found to be some of the most stable parameters when dealing with morphological classification of active galaxies, in the case of both early- and late-types, in combination with CABR. CCON and M20 can be used for classifying both early- and late-type AGN hosts up to 10\%\,-\,25\% of AGN contribution to the total optical light, but not higher. We suggest avoiding SMOOTH parameter when classifying active galaxies, or to use it at least in combination with other parameters.\\
\item In line with previous, morphological diagrams based on the combination of ASYM, CABR, and GINI parameters are
much more stable, and even when dealing with 50\% and 75\% of AGN contribution, a significant fraction of $\sim$\,50\% of early- and late-type galaxies can still be classified. This is not the case when either CCON and/or M20 parameters are included, when $>$\,40\% contamination between early- and late-types has been obtained for 25\% of AGN contribution, and $>$\,60\% in case of 50\% and 75\% of AGN contribution to the total optical light.\\
\end{itemize}

\section*{Acknowledgments}
We thank the anonymous referee for accepting to review this paper, giving us constructive comments that improved our manuscript. TGW acknowledges the support from Bule Hora University under the Ministry of Science and Higher Education. TGW, MP, and ZBA acknowledge financial support from the Ethiopian Space Science and Technology Institute (ESSTI) under the Ethiopian Ministry of Innovation and Technology (MInT). TGW, JM, and MP acknowledge support given through the grant CSIC I-COOP 2017, COOPA20168. MP, JM, JP, and IMP acknowledge the support from the Spanish Ministerio de Ciencia e Innovaci\'on - Agencia Estatal de Investigaci\'on through projects PID2019-106027GB-C41 and AYA2016-76682C3-1-P, and the State Agency for Research of the Spanish MCIU through the Center of Excellence Severo Ochoa award to the Instituto de Astrof\'isica de Andaluc\'ia (SEV-2017-0709). In this work, we made use of Virtual Observatory Tool for OPerations on Catalogues And Tables (TOPCAT).
\section*{DATA AVAILABILITY}
In support of this study, no new data were generated or analyzed. The data used in this article can be obtained from the public sources cited in the article (or references therein).


\bsp    
\end{document}